\documentclass[%
 reprint,
 amsmath,amssymb,
 aps,
 showkeys,
]{revtex4-2}

\usepackage{comment}
\usepackage{graphicx}
\usepackage{dcolumn}
\usepackage{xfrac}
\usepackage{bm}
\usepackage{braket}
\usepackage{BOONDOX-calo}
\usepackage{hyperref}
\usepackage{multirow}
\usepackage{xcolor}
\hypersetup{colorlinks=true, linkcolor=[rgb]{0.00,0.00,1.00}, citecolor=[rgb]{0.00,0.00,1.00}, urlcolor=[rgb]{0.00,0.00,1.00}}

\newcommand{\mbold}[1]{\boldsymbol{\mathrm{#1}}}

\newcommand{\mrm}[1]{\mathrm{#1}}

\newcommand{\beginsupplement}{%
        \setcounter{table}{0}
        \renewcommand{\thetable}{S\arabic{table}}%
        \setcounter{figure}{0}
        \renewcommand{\thefigure}{S\arabic{figure}}%
     }
\graphicspath{{Figures/}}
\newcolumntype{x}[1]{>{\centering\arraybackslash\hspace{0pt}}p{#1}}

\newcommand{\appendixSI}{Appendix}
\begin{document}
\preprint{APS/123-QED}
\title{Trends in the electronic structure of borophene polymorphs}
\newcommand{\lsi}{LSI, CNRS, CEA/DRF/IRAMIS, \'Ecole Polytechnique, Institut Polytechnique de Paris, F-91120 Palaiseau, France}
\newcommand{\etsf}{European Theoretical Spectroscopy Facility (ETSF)}

\author{Alam Osorio}
\affiliation{\lsi}
\affiliation{\etsf}

\author{Lucia Reining}
\affiliation{\lsi}
\affiliation{\etsf}

\author{Francesco Sottile}
\affiliation{\lsi}
\affiliation{\etsf}

\date{\today}

\begin{abstract}

Borophene is a two-dimensional material made out of boron atoms only. It exhibits polymorphism and different allotropes
can be studied in terms of a rigid electronic structure, where only the occupation of the states change with the respect
to the number of electrons available in the system (self-doping). In this work we selected a set of representative
borophene polymorphs ($\delta_3$, $\delta_5$, $\delta_6$, $\beta_{12}$ $\alpha_1$, $\alpha'$, $\alpha'$-Bilayer) and studied the
shared features of their electronic structures and the limitations of this model. Our work revealed the appearance of
defect-like states in some polymorphs when related to a parent rigid electronic structure, and bonding/antibonding
monolayer-like states in the $\alpha'$-Bilayer.
Moreover, we show how the buckling of $\delta_6$ and $\alpha'$ can act as a tuning parameter, enabling semimetallicity,
Dirac cones, and nesting of the Fermi surface.
In light of their promises for exotic but also useful behavior, we expect our work to 
foster the interest in larger and more complex borophene structures. 

\end{abstract}

\maketitle


\section{Introduction}
Two-dimensional materials are of interest for numerous applications, including fields such as
plasmonics, transparent conductors, superconductivity or energy storage \cite{10.1038/nenergy.2017.89,
10.1557/s43578-022-00721-z,10.1038/s41578-019-0136-x,10.1038/natrevmats.2016.55,10.1038/natrevmats.2016.42,
10.1038/natrevmats.2016.52,10.1002/adma.202200574}. In relation to these applications or to fundamental questions, different families of materials have
met strong interest. These range from perovskites \cite{10.1038/s41563-021-01029-9}, transition metal
dichalcogenides (TMDs) \cite{10.1038/natrevmats.2017.33}, and graphene and related materials
\cite{10.1038/nature11458, 10.1103/PhysRevB.78.045405, 10.1007/s40820-020-00464-8}, to more specific and less studied, but nevertheless not less interesting, 
families. A particularly intriguing one is constituted by boron-only two-dimensional (2D) structures, called borophene, in analogy to graphene
\cite{10.1038/ncomms4113}. However, while graphene is precisely defined as the widely studied carbon-only
hexagonal structure with its $sp^{2}$ bonds and a $p_{z}$ orbital, borophene refers to many theoretical and/or
experimentally found polymorphs \cite{10.1021/nl3004754, Wang2024}. This polymorphism is due to the fact that, contrary to carbon, boron has only 3 valence electrons, which leads to  frustration with respect to the perfect
$sp^{2}+p_{z}$ configuration of the hexagonal graphene structure\cite{10.1103/PhysRevB.82.115412}. This problem is overcome by polymorphs with additional or missing atoms in various places.

Numerous theoretical studies of borophene can be found in the literature, which fall essentially into two groups: on one side, a big effort is devoted to 
 finding stable polymorphs, mainly studying formation energies and phonon spectra
\cite{10.1021/nl3004754,10.1021/nn302696v,
10.1021/jp305545z,10.1021/acs.jpclett.7b00891}; on the other hand, many studies focus on
specific properties of certain polymorphs and their possible applications, for example
in superconductivity \cite{Yan2021,10.1021/acs.nanolett.6b00070,10.1021/acs.jpcc.8b03108,10.1103/PhysRevB.98.134514,10.1063/1.4953775,10.1063/1.4963179}, supercapacitance \cite{10.1016/j.apsusc.2021.150154,Pal2025}, hydrogen storage
\cite{10.1039/C7RA12512G,10.1021/jp9077079}, plamonics \cite{10.1103/PhysRevLett.125.116802,
10.1021/jacs.7b10329,10.1021/acs.nanolett.9b04789}, and transparent
conductors\cite{10.1021/acsomega.7b01232,10.1021/acs.jpcc.7b10197}.
In spite of the growing interest in this materials family, only few
highly studied polymorphs have been synthesized: the first experimental realization of a borophene monolayer
was achieved by Mannix \textit{et. al.}, in 2015, obtaining the triangular buckled structure known as $\delta_{6}$
on top of a silver substrate\cite{10.1126/science.aad1080}. Other
polymorphs were then synthesized by different groups on top of metallic substrates: $\beta_{12}$ and
$\chi_{3}$\cite{10.1038/nchem.2491}, $\delta_{3}$ 
\cite{https://doi.org/10.1016/j.scib.2018.02.006}, and the more recent $\alpha^{\prime}$-Bilayer
\cite{10.1038/s41563-021-01084-2}. Two structures, $\beta_{12}$ and
$\chi_{3}$, and intermediate phases, were reported experimentally in their free-standing form (2019)
\cite{https://doi.org/10.1002/adma.201900353}.
Over the last years, also more complex
structures were synthesized, such as superstructures with line defects
\cite{10.1038/s41563-018-0134-1}, concentric superlattices \cite{10.1021/acs.nanolett.9b04798},
large bilayers \cite{10.1038/s41557-021-00813-z}, hydrogenated borophene structures known as
borophanes \cite{doi:10.1126/science.abg1874}, and, more recently, also nanoribbons \cite{Li2024}.

The main objective of the present work is to shed additional light on the electronic properties of borophene polymorhps, to examine
how these properties are related to the atomic structure, and to discuss when and how one can exploit 
the differences and similarities to discern patterns and tune parameters to explore exotic behaviour. To this aim, we mainly concentrate 
 on the structures from Refs. \cite{10.1126/science.aad1080,10.1038/nchem.2491,
https://doi.org/10.1016/j.scib.2018.02.006,10.1038/s41563-021-01084-2,https://doi.org/10.1002/adma.201900353},
as they contain only few atoms per unit cell, which eases the analysis. 

The article is organized as follows: after this introduction, a strategy to compare the borophene polymorphs is proposed in Section \ref{sec:comp}. The atomic structure of the 
polymorphs examined in the present work is presented in Section \ref{sec:atomic}. The results and discussion of the electronic
properties, their relation with the atomic structure, and the comparative analysis is carried out in Section \ref{sec:results}, 
for the three main categories: flat monolayers, buckled monolayers, and bilayers. Conclusions in Section \ref{sec:conclusions},
Computational Details in Section \ref{sec:compdet} and \appendixSI\ in Section \ref{sec:supp} complete the work.  

\section{Comparing Borophene Polymorphs}
\label{sec:comp}

\begin{figure*}
\centering
\includegraphics[width=0.95\textwidth]{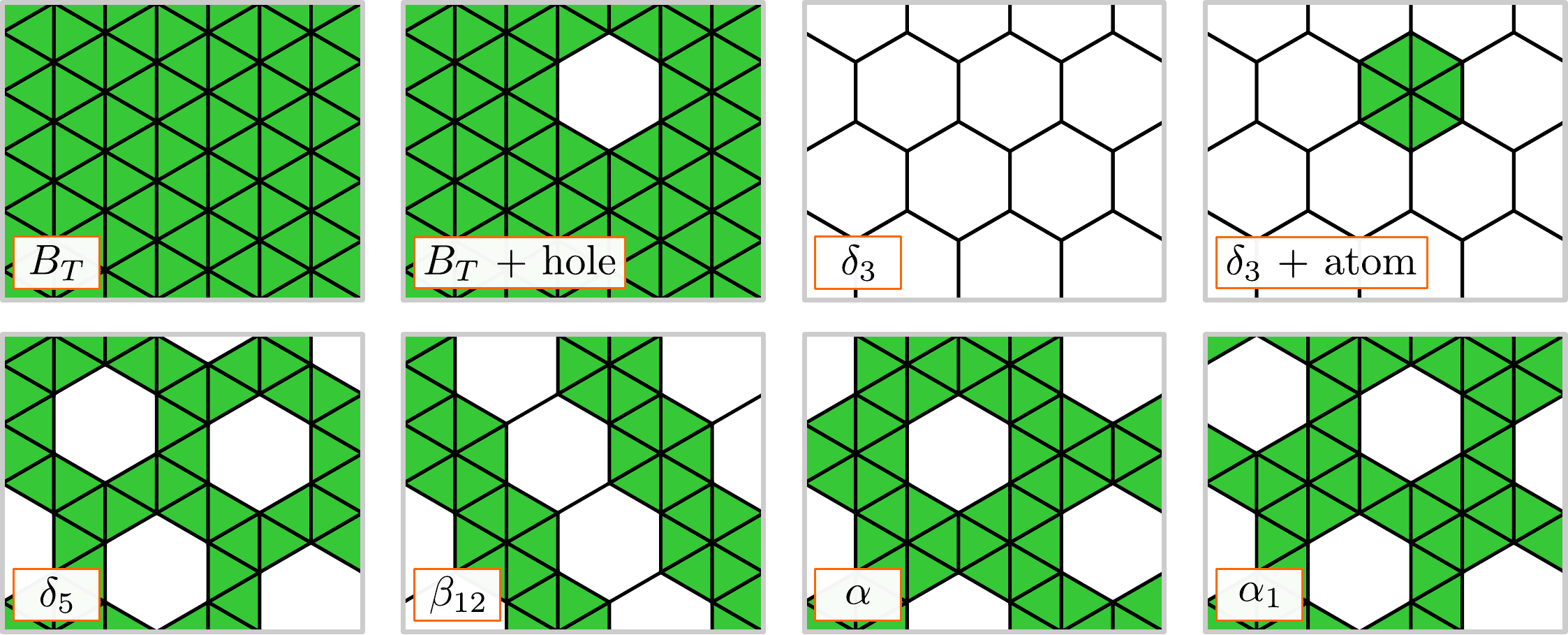}
\caption{\label{fig:DrawingHexagons}Model borophene structures. Each vertex corresponds 
to an atomic position. $B_{T}$ and $\delta_{3}$ (top row) can be used as parent structures of other 
allotropes through the creation of holes (empty hexagons in white) and the addition of atoms (filled 
hexagons in green). Both mechanisms can be used conceptually to construct the child structures 
(lower row) from $B_{T}$, and in certain cases, from $\delta_{3}$. Moreover, in an $sp^{2}$ bonding configuration in-plane 
(like graphene), $\delta_{3}$ is electron-deficient, while the $B_{T}$ lattice has an excess of electrons, 
thus the children structures can be seen as a competition of the two lattices, where the balance is 
achieved for the $\alpha$ structure\cite{10.1103/PhysRevLett.99.115501}.}
\end{figure*}

Different flat monolayer polymorphs can be described by
starting from a common model structure: a periodic array of hexagonal building units in which 
every vertex describes an atomic position in-plane.
This structure corresponds to a polymorph that has been created experimentally, known as $\delta_{3}$ 
(see Fig. \ref{fig:DrawingHexagons}, third structure in the top row). It is the equivalent of 
graphene, but constructed with boron rather than carbon atoms. The $\beta_{12}$ structure, shown as second in the bottom row of Fig. \ref{fig:DrawingHexagons}, relies on this
 hexagonal pattern, but some of the hexagons contain a boron atom in the center. One may think of 
$\delta_{3}$ as a parent structure to which one adds periodically an atom in order to obtain the child structure $\beta_{12}$.

One can create many other polymorphs starting from $\delta_{3}$, for example the $\alpha$ 
monolayer (Fig. \ref{fig:DrawingHexagons}, third in the bottom row). 
However, the empty hexagonal building units are not always
compatible with the periodicity of more complex structures (see for example $\delta_{5}$ and
$\alpha_{1}$ in Fig. \ref{fig:DrawingHexagons}). Hence, it is convenient to define another parent
structure, where all hexagons are filled by a boron atom in the center: we will refer to this model system
as $B_{T}$ (see Fig. \ref{fig:DrawingHexagons}, top left). Starting from $B_{T}$, a second conceptual mechanism
to build other polymorphs is to remove atoms from the full triangular mesh.
This gives more flexibility than starting from $\delta_3$, since any of the other polymorphs can be obtained in this way. However, it is useful to keep both parent structures for the following discussions. 
Moreover, in a graphene-like $sp^{2}$ bonding configuration in-plane $\delta_{3}$ is electron-deficient, while the $B_{T}$ lattice has an excess of electrons, and therefore 
the child structures can be seen as a competition of the two lattices, where the balance is 
achieved for the $\alpha$ structure\cite{10.1103/PhysRevLett.99.115501}.
Note that when comparing the different structures, we will suppose that bond lengths remain unchanged, unless stated differently, which is justified by the fact that some structures are in any case hypothetical, by the fact that here we are interested in global trends, and by results concerning changes due to bond lengths that will be discussed in Section \ref{sec:results}.

The ${B_{T}}$ structure has a characteristic atomic density $\rho_{B_T}$, defined as the number of boron atoms $N_{B_{T}}$ per unit area $A$,
\begin{equation}
\rho_{B_T}=\frac{N_{B_T}}{A}.
\label{eq:atomicdensity_bt}
\end{equation}

Creating holes in the model system to construct another polymorph $P$ that is described by a different number of boron
atoms $N_{B_P}$ results in a different atomic density,
$\frac{N_{B_{P}}}{A}$, in the
child structure.

It is common practice \cite{10.1103/PhysRevLett.99.115501,PhysRevB.80.134113,10.1103/PhysRevB.82.115412,
10.1038/s41467-023-37442-8,10.1021/nl3004754,10.1021/jp305545z,10.1021/nn302696v,doi:10.1021/acs.jpclett.0c01359,
10.1039/D2NH00518B} to define the change with respect to $B_{T}$ as 
\begin{equation}
\eta_{P} = \frac{\rho_{B_{T}}-\rho_{B_{P}}}{\rho_{B_{T}}}= \frac{N_{B_{T}}-N_{B_{P}}}{N_{B_{T}}},
\label{eq:eta_Equation}
\end{equation}

\noindent where $B_T$ and polymorph $P$ are described in unit cells of same area $A$.  In other words, $\eta_{P}$ 
corresponds to the number of holes per unit area that exist in a given polymorph $P$ with
respect to the triangular mesh $B_T$. For example, $\delta_{3}$ in its primitive unit
cell has $N_{\delta_{3}}=2$, while for the same unit cell $N_{B_{T}}=3$, and therefore 
$\eta_{\delta_{3}}=\sfrac{1}{3}$.
Moreover, one can use this parameter $\eta_{P}$ to express the density of valence electrons in a given polymorph, with 3 valence electrons
per boron atom,  as 
\begin{equation}
\rho^{P}_{\mathrm{el}} = 3\rho_{B_T}(1-\eta_{P}).
\label{eq:rhoEl_withEta}
\end{equation}

The parameter $\eta_{P}$ is often used as a basic descriptor of the different polymorphs, and some general
observations in borophene have been linked to it: perhaps the most notorious examples are the prediction
of polymorphism, with 
configurations of low energy occurring in a range of $\eta_{P}$ around 0.1-0.15 \cite{10.1021/nl3004754}, and the introduction of a concept known as self-doping\cite{PhysRevB.80.134113}, which will be discussed in the following.

\subsubsection{The self-doping picture}

The search of stable borophene polymorphs is an area of intense research, and the problem is
tackled from different perspectives: phonon spectra\cite{10.1021/nn302696v,10.1021/acs.jpclett.7b00891,
10.1021/jp8052357,10.1021/acs.jpcc.9b03447,10.1021/jp8052357}, formation energies\cite{10.1021/nl3004754,
10.1021/jp305545z,https://doi.org/10.1002/anie.201705459,doi:10.1021/acs.jpclett.0c01359}, temperature-dependent
dynamics\cite{doi:10.1021/acs.jpclett.0c01359,10.1021/acs.jpclett.7b00891,10.1039/D2NH00518B} and electronic
stability\cite{10.1103/PhysRevLett.99.115501,10.1038/s41467-023-37442-8}. Among the latter, the
works of Tang and Ismail-Beigi describe the lowest-energy electronic configurations of the polymorphs in
terms of the bonding and antibonding character of the states\cite{10.1103/PhysRevLett.99.115501,PhysRevB.80.134113}:
in these studies $\eta_{P}$ plays a key role as an indicator of stability. 

In Ref. \cite{10.1103/PhysRevLett.99.115501}, bonding and antibonding in-plane states of different polymorphs
form a gap in the computed density of states. Polymorphs in which bonding states are
not fully occupied are electron deficient, hence electronically unstable;
The more electronically stable configurations are those in which
the bonding states in-plane are fully occupied,
while the in-plane antibonding ones are empty. This condition is fulfilled in the $\alpha$ structure, where $\eta_{\alpha}=1/9$ (see Fig. \ref{fig:DrawingHexagons}).
Polymorphs with $\eta_{P}<\sfrac{1}{9}$ have an excess of electrons with respect
to the optimal $\eta_{\alpha}$ configuration.
In Ref. \cite{10.1103/PhysRevLett.99.115501}, $\eta_{P}$ was therefore proposed as an indicator for the optimal filling of the
states and deviations from optimal filling. Moreover, it was suggested that upon the addition or removal of atoms, the
bonding and antibonding states in the electronic structure are not changed, but just filled differently according
to the number of electrons by shifting the Fermi level. 
This hypothesis for the electronic structure was later formulated
by Tang and Ismail-Beigi in terms of a concept known as ``self-doping''\cite{PhysRevB.80.134113}:  borophene is supposed to exhibit a rigid electronic structure, in which the change of the atomic structure can be simply linked to the
change of the atomic density $\rho_{B_{P}}$, and therefore the vacancy ratio, $\eta_{P}$, which in turn can be
directly linked to the number of electrons (Eq. \ref{eq:rhoEl_withEta}) and hence the filling of the states. In
other words, the electronic structure of different polymorphs can be estimated by using a fixed borophene
electronic structure (for example of the parent structure $B_{T}$), and the vacancy ratio $\eta_{P}$.
Numerous first-principles calculations
\cite{10.1021/nl3004754,doi:10.1021/acs.jpclett.0c01359,https://doi.org/10.1002/anie.201705459,10.1021/jp305545z,
10.1039/D2NH00518B}
have shown $\eta_{P}$ to be good indicator for the stability of borophene, and thus it has been largely adopted in literature.

In the present work we aim at a more detailed understanding of the role of the atomic structure in determining the electronic properties of 
various borophene polymorphs. We use $\eta_{P}$ and the self-doping mechanism as a starting point for understanding
the similarities and differences of their electronic structures, and we highlight the validity as well as the limitations of this approach, which point to a need for
full ab initio calculations for a series of interesting questions.

\section{\label{sec:atomic}Atomic structures}
\label{sec:atomic}

\begin{figure*}
\centering
\includegraphics[width=0.95\textwidth]{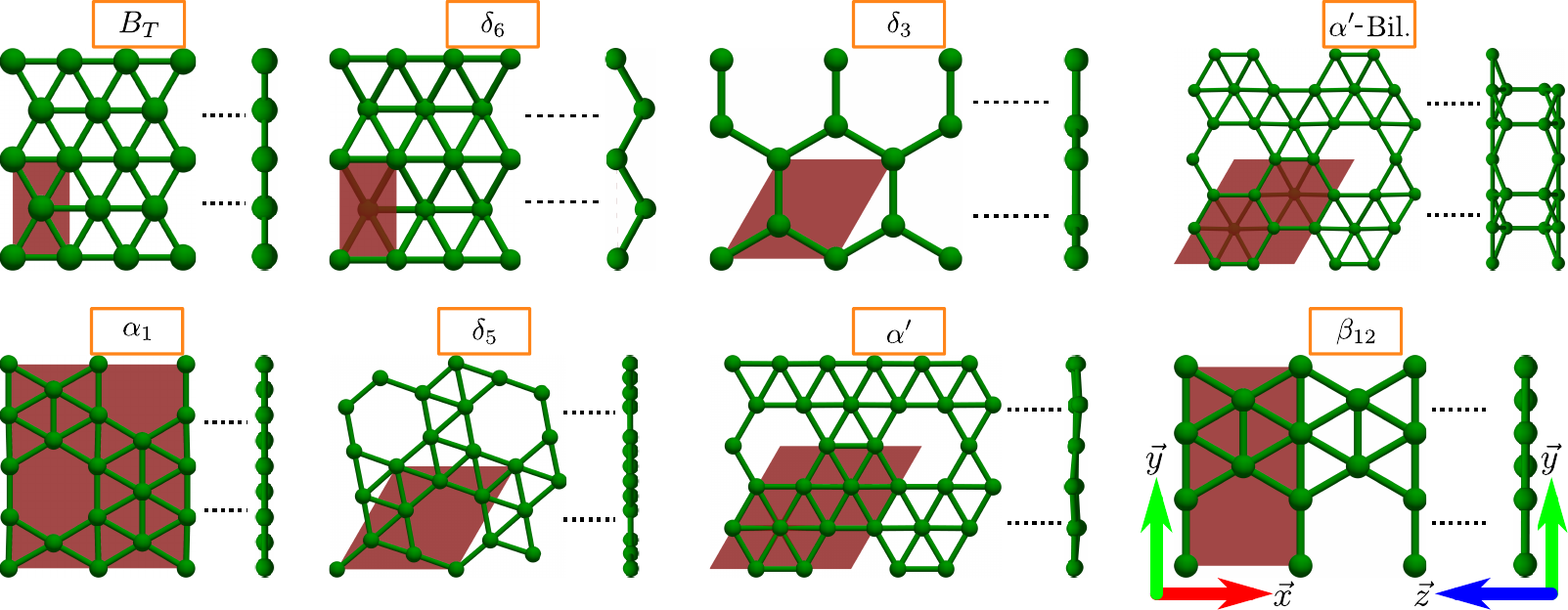}
\caption{\label{fig:AtomicStructures_Polymorphs}Atomic structure of borophene polymorphs, including flat, buckled and bilayer structures from Refs. \cite{10.1021/jp066719w, 10.1021/acs.jpclett.7b00891, 10.1021/nn302696v, 10.1126/science.aad1080, 10.1103/PhysRevLett.125.116802, 10.1038/s41563-021-01084-2}. As all polymorphs are quasi-two-dimensional we present their projection in the $xy$-plane, accompanied by a projection of the structure in the $yz$-plane. In our calculations we have used the unit cells highlighted in brown, unless otherwise specified.}
\end{figure*}

For our study, we selected different polymorphs based on their structural differences and relevance for the
theoretical and experimental results found in literature. We included five flat monolayers: $B_T$,
$\delta_{3}$, $\beta_{12}$, $\alpha_{1}$ and $\delta_5$, two buckled monolayers: $\alpha^{\prime}$ and $\delta_6$,
and the AA-stacked $\alpha^{\prime}$-bilayer, from Refs. \cite{10.1021/jp066719w, 10.1021/acs.jpclett.7b00891,
10.1021/nn302696v, 10.1126/science.aad1080, 10.1103/PhysRevLett.125.116802, 10.1038/s41563-021-01084-2}. Atomic
structures are shown in Figure \ref{fig:AtomicStructures_Polymorphs}. In the following, we describe the unit cells of the selected
polymorphs in terms of symmetry, atomic positions (Wyckoff sites), lattice parameters,
number of atoms per unit cell and vacancy ratio.

\paragraph*{$\mbold{B_T}$:}This structure is a model system, flat and regular, that is used only for
comparison with other structures. Indeed, this polymorph buckles upon the full optimization of the
atomic positions, which gives rise to the $\delta_6$ structure. The lattice parameters 
and atomic positions in the ideal unbuckled $B_{T}$ model have to be imposed. Some constrained calculations can be found in the literature
\cite{10.1021/jp8052357, 10.1103/PhysRevLett.99.115501}, suggesting equal bond lengths. Here, we will work with one of the two different
bond lengths that occur in the $\delta_{6}$ structure, namely,
 1.68 $\mathrm{\AA}$. This choice has no significant impact, as can be seen in  
Fig. \ref{fig:BT_diffBonds_WStructure} in the \appendixSI.

\paragraph*{$\mbold{\delta_3}$:}The graphene-like structure belongs to the group of the synthesized borophene
polymorphs \cite{https://doi.org/10.1016/j.scib.2018.02.006}. It contains two atoms per unit cell in a hexagonal
system with symmetry \textit{P6/mmm} at the positions \textit{2c}. The atomic positions and lattice parameters
were obtained from Ref. \cite{10.1021/jp066719w}, where the in-plane parameters $a=b=2.91$
$\mathrm{\AA}$ were calculated with DFT-GGA using the Perdew-Wang exchange-correlation functional\cite{10.1103/PhysRevB.45.13244}.

\paragraph*{$\mbold{\beta_{12}}$:} This is a popular polymorph that has been synthesized and used
as a prototype system due to its high stability and promising properties regarding transparent conductors, superconductivity
and plasmonics, among others \cite{10.1038/nchem.2491, 10.1103/PhysRevB.94.041408, https://doi.org/10.1002/adma.201900353,
10.1103/PhysRevLett.125.116802, 10.1021/acs.jpcc.7b10197, 10.1021/acs.nanolett.6b00070}. It contains five atoms per
unit cell in a structure with symmetry \textit{Pmmm}, with positions \textit{2m},  \textit{2o} and \textit{1e},
the latter being in the center of the hexagonal building unit. The in-plane dimensions for this structure were taken
from the calculations of Ref. \cite{10.1103/PhysRevLett.125.116802}: $a=2.92$ \AA{} and $b=5.06$ \AA. These parameters are
 in good agreement with the
theoretical (GGA-PBE) and experimental parameters of $\beta_{12}$ on Ag(111) from Refs. \cite{10.1038/nchem.2491,
10.1103/PhysRevB.94.041408}.

\paragraph*{$\mbold{\alpha_1}$:}The theoretical research concerning the structural features of borophene, mainly based
on formation energies and phonons, lead to a structure known as $\alpha_{1}$. 
It has been recently predicted to appear under low boron concentration
and low temperatures \cite{10.1039/D2NH00518B}. This polymorph contains 14 atoms per unit cell in a structure with
\textit{Cmmm} symmetry, localized at the positions \textit{4i}, \textit{8p} and \textit{2b}. The lattice parameters
($a=5.84$ \AA, $b=6.74$ \AA) and atomic positions were obtained in Ref. \cite{10.1021/nn302696v} from DFT
calculations using the exchange-correlation functionals PBE and PBE0 \cite{10.1063/1.478522}.

\paragraph*{$\mbold{\delta_5}$:}Different structures have emerged as a result of the search for stable polymorphs
with specific properties, among which we highlight the case of $\delta_5$, the only polymorph 
(besides on of the parent structure, $\delta_3$) here presented without any
six-coordinated atom. It was predicted to show linearly dispersing bands 
crossing at the $\mathrm{K}$ point, similarly to the Dirac cone in graphene, above the Fermi energy 
\cite{10.1021/jp305545z, 10.1021/acs.jpclett.7b00891}. The unit cell
has symmetry \textit{P6/m} with lattice parameters $a=b=4.47$ $\mathrm{\AA}$, with six atoms at the sites
\textit{6j}, as obtained from Ref. \cite{10.1021/acs.jpclett.7b00891} via DFT-GGA calculations using the PBE
exchange-correlation functional.

\paragraph*{$\mbold{\alpha^{\prime}}$:}In terms of formation energy and electronic stability, one of the most
investigated structures is the planar $\alpha$ polymorph \cite{10.1103/PhysRevLett.99.115501,
10.1103/PhysRevB.82.115412, 10.1021/nl3004754, 10.1103/PhysRevB.77.041402}. It was, however, found to be unstable
and it was predicted to show a slight distortion out-of-plane. The resulting polymorph was called $\alpha^{\prime}$
\cite{10.1021/nn302696v}. The unit cell of this polymorph has symmetry \textit{P-3m1}, with 8 atoms at the positions
\textit{6g} and \textit{2d}. For this case, the atomic positions and lattice parameters were optimized in the present
work using DFT-PBE, yielding lattice parameters
 $a=b=5.058$ $\mathrm{\AA}$,with a buckling off-plane of
the two atoms by $\pm 0.085$ $\mathrm{\AA}$.  The optimized structure can be compared with Refs. \cite{10.1021/nn302696v,
10.1103/PhysRevB.93.165434}. In Ref. \cite{10.1021/nn302696v}, $a=5.046$ \AA{} and $b=5.044$ \AA{} (DFT-PBE) with
buckling of $\pm 0.17$ $\mathrm{\AA}$ (atomic positions optimized with PBE0 on top of the PBE-optimized
lattice parameter), while in Ref. \cite{10.1103/PhysRevB.93.165434} $a=b=5.10$ $\mathrm{\AA}$ with a
buckling of $\pm 0.14$ $\mathrm{\AA}$ (DFT-PBE). Despite the differences in the atomic structure, particularly with
respect to buckling, the electronic structure is very similar (See \appendixSI, Fig. \ref{fig:Balpha_Bands_LDA_PBE}).

\paragraph*{$\mbold{\delta_6}$:}The flat triangular structure $B_T$
has been shown to be thermodynamically unstable, and it has been predicted to buckle, leading to the so-called
$\delta_6$ polymorph \cite{PhysRevB.72.045434, 10.1021/jp8052357}. This system has symmetry \textit{Pmmn} and contains
two atoms per unit cell at positions \textit{2a}. The in-plane lattice parameters are $a=1.62$ and $b=2.87$
$\mathrm{\AA}$, with a buckling height of $0.80$ $\mathrm{\AA}$. These were obtained in Ref. \cite{10.1126/science.aad1080}
(DFT-PBE), in good agreement with the experimental results on Ag(111) in the same reference. 

\paragraph*{$\mbold{\alpha^{\prime}}$-Bilayer:}We selected this polymorph for easy comparison with the
$\alpha^{\prime}$-monolayer. The unit cell, containing 16 atoms,  
has symmetry \textit{P6/mmm}, and the atoms are at the positions
\textit{12n} and \textit{4h}. For this polymorph we optimized the lattice parameters and atomic positions, obtaining $a=b=5.70$ $\mathrm{\AA}$, with distances among the
bottom layer and top layer of $1.80$ \AA{} and $3.10$ \AA, for the inner and outer atoms of the bilayer, respectively. These values are in excellent agreement with the reference structure from Ref. \cite{10.1038/s41563-021-01084-2} calculated
with DFT-PBE, and their experimental measurements on Ag(111).

\noindent We summarize the atomic structure of the different polymorphs as used in this work in Table
\ref{tab:StructuralParameteresofAtomicPositions}, whose columns illustrate vacancy ratio with respect 
to $B_T$, symmetry, lattice parameters, and Wyckoff sites.

\begin{table}[h]
\begin{tabular}{c|c|c|c|c|c}

 & $\eta_{P}$ & Symmetry & $a$ & $b$ & Wyckoff sites \\ \hline
  $B_T$             & 0            & \textit{P6/mmm} & 1.68 & 1.68 & \textit{1a} $(0.00, 0.00, 0.00)$ \\ \hline
$\delta_3$        & $\sfrac{1}{3}$ & \textit{P6/mmm} & 2.91 & 2.91 & \textit{2c} $(0.33, 0.66, 0.00)$ \\ \hline
$\beta_{12}$      & $\sfrac{1}{6}$ & \textit{Pmmm}   & 2.92 & 5.06 & \begin{tabular}[c]{@{}c@{}}\textit{2m} $(0.00, 0.83, 0.00)$\\ \textit{2o} $(0.50, 0.33, 0.00)$\\ \textit{1e} $(0.00, 0.50, 0.00)$\end{tabular} \\ \hline
$\alpha_1$        & $\sfrac{1}{8}$ & \textit{Cmmm}   & 5.84 & 6.67 & \begin{tabular}[c]{@{}c@{}}\textit{4i} $(0.00, 0.75, 0.00)$\\ \textit{8p} $(0.75, 0.12, 0.00)$\\ \textit{2b} $(0.50, 0.00, 0.00)$\end{tabular} \\ \hline
$\delta_5$        & $\sfrac{1}{7}$ & \textit{P6/m}   & 4.47 & 4.47 & \textit{6j} $(0.57, 0.72, 0.00)$\\ \hline
$\alpha^{\prime}$ & $\sfrac{1}{9}$ & \textit{P-3m1}  & 5.06 & 5.06 & \begin{tabular}[c]{@{}c@{}}\textit{6g} $(0.33, 0.00, 0.00)$\\ \textit{2d} $(0.33, 0.66, -0.00)^{\ast}$\end{tabular} \\ \hline
$\delta_6$        & ---            & \textit{Pmmn}   & 1.62 & 2.87 & \textit{2a} $(0.25, 0.25, -0.02)$\\ \hline
Bil.              & ---            & \textit{P6/mmm} & 5.70 & 5.70 & \begin{tabular}[c]{@{}c@{}}\textit{12n} $(0.33, 0.00, -0.07)$\\ \textit{4h} $(0.33, 0.66, -0.04)$\end{tabular}
\end{tabular}

\caption[]{Structural parameters of selected borophene polymorphs as used in this work. The atomic positions are reported in reduced coordinates (rounded to two decimals), with $c=19.05$ $\mathrm{\AA}$ ($c=21.17$ $\mathrm{\AA}$ for the $\alpha^{\prime}$-Bilayer). $^{\ast}$ Note that the out-of-plane distortion of $-0.0045c$ in $\alpha^{\prime}$ is smaller than the rounding in the table.}
\label{tab:StructuralParameteresofAtomicPositions}
\end{table}

It is worth remarking that the literature concerning the search of stable polymorphs is extensive and
goes far beyond the structures here presented: for instance, one can find other bilayer
structures \cite{10.1038/s41557-021-00813-z,10.1038/s41598-022-18076-0}, borophene with large
holes\cite{doi:10.1021/acs.jpclett.0c01359, 10.1039/D2NH00518B}, intermixed
superstructures\cite{10.1038/s41563-018-0134-1,10.1021/acs.nanolett.9b04798},
polymorphs with Dirac cones\cite{10.1103/PhysRevLett.112.085502,10.1021/acs.jpclett.7b00891} and 
Kagome lattices\cite{gao2023bilayer}, among others, all involving a single atomic species: boron. 
Therefore, borophene constitutes an excellent playground to study how the atomic
structure determines the properties of the system.

The electronic structures presented in the following were obtained by DFT-LDA calculations as described in Sec. \ref{sec:compdet}.

\section{\label{sec:results}Results and discussion}

\begin{figure*}
\centering
\includegraphics[width=0.95\textwidth]{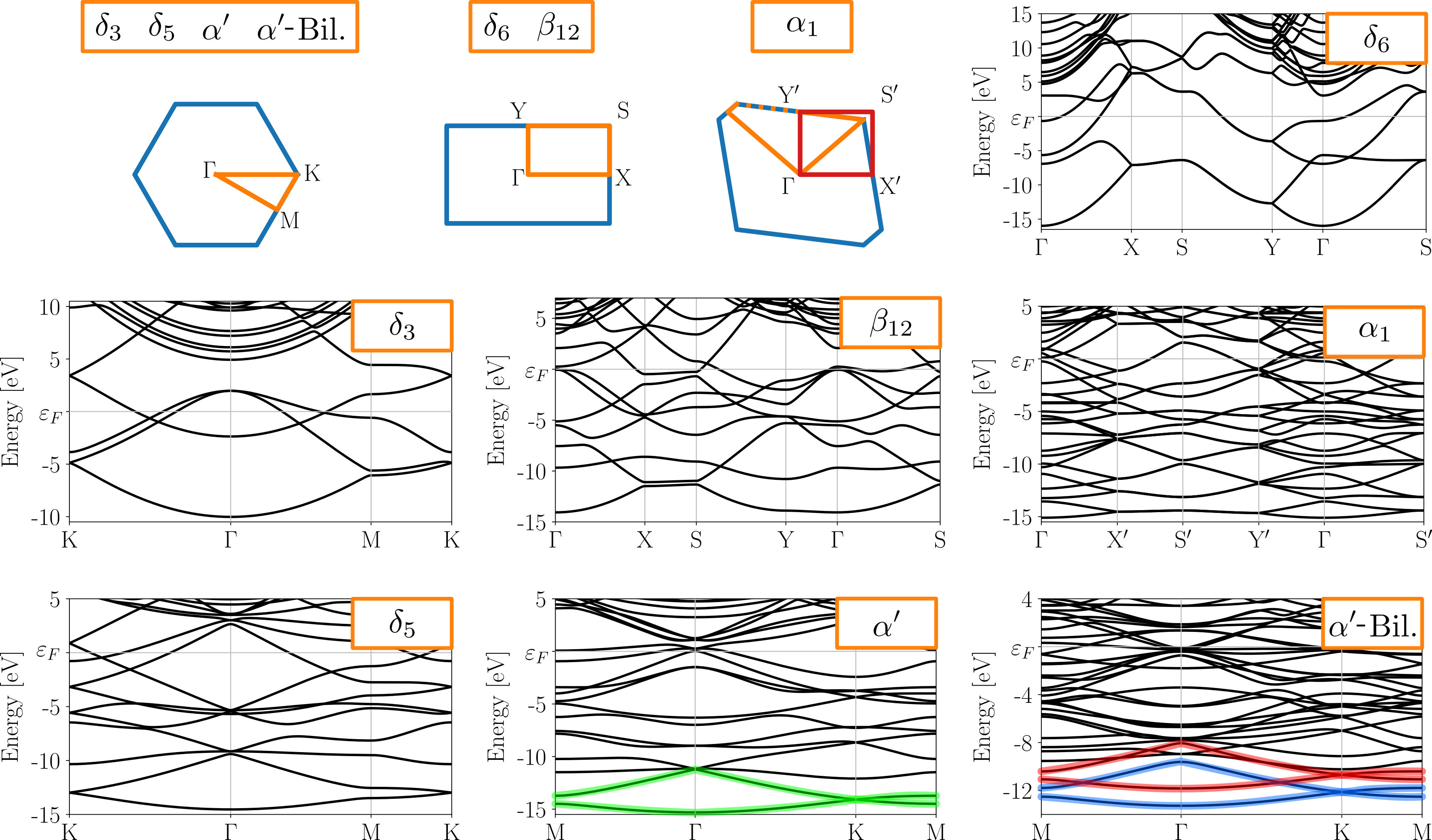}
\caption{\label{fig:BandStructureAllPolymorphs} Brillouin zones (BZs) and band structures of different borophene polymorphs. All the polymorphs here presented are metallic. Our results are in good agreement with previously reported ones \cite{10.1021/acs.jpcc.9b03447, https://doi.org/10.1016/j.ssc.2018.08.003, 10.1021/acs.jpclett.7b00891, 10.1021/nn302696v, 10.1103/PhysRevLett.125.116802, 10.1038/s41563-021-01084-2, 10.1038/nchem.2491, 10.1126/science.aad1080, 10.1103/PhysRevB.98.134514, 10.1039/C6TC00115G}. For $\alpha_1$, of symmetry $Cmmm$, the band structure is not computed along the high symmetry paths of the BZ, but along a non-conventional path including the $\mbold{k}$-points used in Ref. \cite{10.1021/nn302696v} to facilitate comparison. The two lowest states in the $\alpha^{\prime}$ structure are highlighted for comparison with the two sets of bonding- and antibonding-like states of the $\alpha^{\prime}$-Bilayer.}
\end{figure*}

Our calculated Kohn-Sham band structures of the selected polymorphs  
are summarized in Fig. \ref{fig:BandStructureAllPolymorphs}. All polymorphs are metallic. 
We found good agreement between our LDA results and results reported in the
literature, including those using different functionals \cite{10.1021/acs.jpcc.9b03447,
https://doi.org/10.1016/j.ssc.2018.08.003,10.1021/acs.jpclett.7b00891, 10.1021/nn302696v,
10.1103/PhysRevLett.125.116802,10.1038/s41563-021-01084-2, 10.1038/nchem.2491, 10.1126/science.aad1080, 
10.1103/PhysRevB.98.134514, 10.1039/C6TC00115G}. Only with respect to Ref. \cite{10.1021/nn302696v}, our calculated fermi level in $\alpha_1$ is shifted downwards by 1 eV, while the band structure is very similar. The origin of this
discrepancy is unclear, but we have checked carefully that our calculation yields the correct number of electrons, and it should therefore be trusted. Note that this implies that the opening of the gap upon the use of PBE0 in Ref. \cite{10.1021/nn302696v} does not change the metallic character of the system.

For the following comparisons, we consider three conceptual ``engineering mechanism'' with
respect to a parent structure: 1) via creation of defects; either creating vacancies from $B_T$
($\overset{\circ}{\longrightarrow}$), or adding interstitial atoms into the hexagonal sites of $\delta_3$
($\overset{\bullet}{\longrightarrow}$), 2) buckling ($\overset{\wedge}{\longrightarrow}$)
as in $\alpha^{\prime}$ and $\delta_{6}$, and 3) the addition of a second layer on top of a monolayer
(comparison between $\alpha^{\prime}$ and $\alpha^{\prime}$-Bilayer). 

\subsection{Flat monolayers}\label{section:ElectronicStructureFlatMonolayers}
\subsubsection{Band structures in commensurable unit cells}

All flat monolayers are related by addition or removal of atoms from a common
parent structure (see for example, Refs. \cite{PhysRevB.80.134113, 10.1021/jp305545z,
10.1103/PhysRevLett.99.115501,10.1021/nl3004754, doi:10.1021/acs.jpclett.0c01359,10.1021/acs.jpclett.7b00891, 
10.1039/D2NH00518B}). However, the change of the primitive unit cell from one polymorph to another, and the related change of the Brillouin zone, complicate the 
comparison. 
Therefore, even when polymorphs have similar lattices, they are most often studied separately. Some attempts have been made to group them, for example with respect to symmetry \cite{10.1021/acs.jpclett.7b00891}, bond nature
\cite{10.1038/s41467-023-37442-8}, and occupation of the states \cite{PhysRevB.80.134113}. The
similarities of certain results for different polymorphs make it interesting to investigate further the impact of the structural changes on the electronic properties. In the following, we will concentrate on a comparison of band structures. 

We selected two representative cases: the creation of $\beta_{12}$ through the addition of atoms into
$\delta_{3}$, which we denote as ($\delta_3\overset{\bullet}{\longrightarrow}\beta_{12}$), and the
creation of $\delta_{5}$ through the removal of atoms from $B_{T}$ 
($B_T\overset{\circ}{\longrightarrow}\delta_5$).
In order to compare two structures, we build a commensurable supercell.
Such a comparison is not obvious when the supercells are very large, due to the band folding in the small Brillouin zone. The two cases proposed here are, however, simple enough for a clear analysis. 

\subsubsection*{Choice of the commensurable unit cell}

The more complex structures $\beta_{12}$ and $\delta_5$, when relaxed, show changes of bond-length with respect to the parent structures $\delta_3$ and $B_T$. This difference of bond-length could lead to larger commensurable supercells and also mask the effect of the mere removal or addition of an atom. Therefore, we fix the bond-lengths of parent and child structures to be equal. In the following, we start from the relaxed child structures $\beta_{12}$ ($\delta_5$) and remove (add) atoms, keeping the bond-lengths fixed, to obtain slightly modified parent structures $\delta_3'$ and $B_T'$. These will be used for the comparisons.

\begin{figure*}[!ht]
\centering
\includegraphics[width=0.95\textwidth]{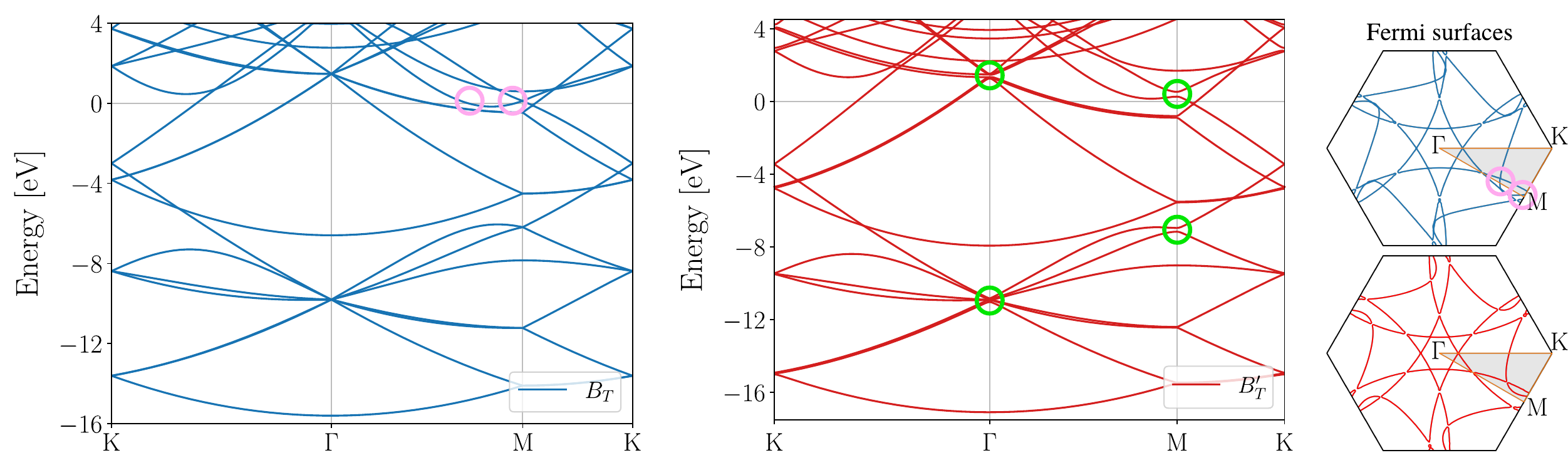}
\caption{\label{fig:BT_BTp_inHex}Band structure of model systems $B_{T}$ (left) and $B_{T}^{\prime}$ (middle panel), where $B_{T}^{\prime}$ is constructed from the atomic positions of $\delta_{5}$.
Right panel: Fermi surfaces of $B_T$ (top) and $B_T^{\prime}$ (bottom). Pink circles highlight crossing of bands with the fermi level in $B_T$ along $\Gamma\to\mrm{M}$. Green circles indicated lifting of degeneracies in $B_T^{\prime}$. }
\end{figure*}

This choice does not impact our analysis, as one can see for the case of $B_T$. Fig. \ref{fig:BT_diffBonds_WStructure} illustrates that the change of the band structure with a uniform scaling of the interatomic distances is visible, but without a qualitative impact, in particular around the fermi level. More importantly, Fig. \ref{fig:BT_BTp_inHex} shows the band structures of the perfectly regular $B_{T}$ and of the $B_{T}'$ structure, which is $B_{T}$-like, but with slightly modified atomic positions. To be precise, we construct $B_{T}'$ from $\delta_{5}$
by taking the atomic position of the latter and adding the missing atom to complete the triangular mesh.
The additional atom was introduced in the center of mass defined by the surrounding atoms.

Moving from $B_T$ to 
$B_{T}^{\prime}$ changes the band width and lifts some degeneracies, as highlighted by the green circles in Fig. \ref{fig:BT_BTp_inHex}. There is also a slight change in the Fermi surface, with a band crossing the Fermi energy in $B_{T}$ along 
$\Gamma\to\mrm{M}$ (pink circles in Fig. \ref{fig:BT_BTp_inHex}), whereas in $B_{T}^{\prime}$  the crossing cannot be seen along $\Gamma\to\mrm{M}$. However, the crossing still occurs, and it is simply found along an only slightly different direction, as one may see from the Fermi surfaces. This justifies the comparisons that we will make in the following by using  $B_{T}^{\prime}$ and also $\delta_{3}^{\prime}$, which is defined along the same lines by using the atomic positions of $\beta_{12}$.

\subsubsection*{Evidence for, and departure from, the self-doping picture}

\begin{figure*}[!ht]
\centering
\includegraphics[width=0.99\textwidth]{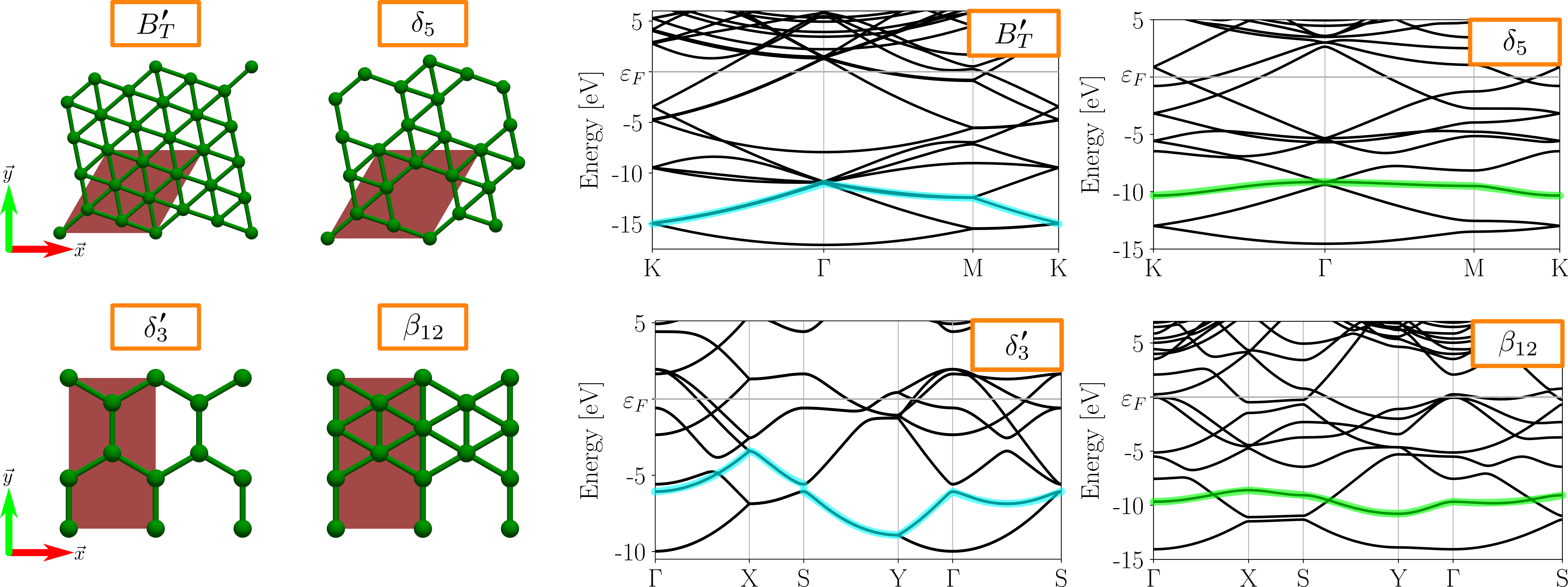}
\caption{\label{fig:BandStructuresSameSettings}Commensurable unit cells of ($B_{T}$ and $\delta_{5}$) and ($\delta_{3}$ and $\beta_{12}$), 
and band structures calculated in the corresponding Brillouin zones. 
A weakly dispersive band is highlighted in green in both $\delta_5$
and $\beta_{12}$. Such weakly dispersive bands are absent in the parent structures $B_{T}$ and $\delta_{3}$. The strongly dispersive parent bands that can be most closely related to these states are highlighted in blue: note the double degeneracy along $\mrm{K}\to\Gamma\to\mrm{M}$ and $\mrm{X}\to\mrm{S}\to\mrm{Y}$ in $B_{T}$ and $\delta_{3}$, respectively.}
\end{figure*}

The bands of the commensurable structures ($B'_T\overset{\circ}{\longrightarrow}\delta_5$)
and ($\delta'_3\overset{\bullet}{\longrightarrow}\beta_{12}$) are grouped in Fig.
\ref{fig:BandStructuresSameSettings}. 
Comparing the respective parent and child polymorphs, one finds that the band structures are overall very similar, including the dispersion of bands far from, as well as close to, the Fermi level. The main difference is a shift of the Fermi level due to the change of number of electrons in the unit cell ($n_{el}^{B'_T}=21$, $n_{el}^{\delta_5}=18$,
$n_{el}^{\delta'_3}=12$, $n_{el}^{\beta_{12}}=15$). This supports the idea to relate different borophene
polymorphs to a reference structure (for example $B'_{T}$ or $\delta'_{3}$), following  
the self-doping picture\cite{PhysRevB.80.134113}, as mentioned in the Introduction.

A closer look at Fig. \ref{fig:BandStructuresSameSettings} shows an interesting fact, which is highlighted in the Figure: as expected, the addition or removal of atoms with respect to the parent structure lifts degeneracies. For some states the effect is so strong that they form an almost flat band. Such flat bands indicate charge localization, and they are particularly interesting in the context of strong correlation when they may occur at the Fermi level. This motivates further investigation below. 

\subsubsection{Nature of the electronic states}

\begin{figure}[!ht]
\centering
\includegraphics[width=0.45\textwidth]{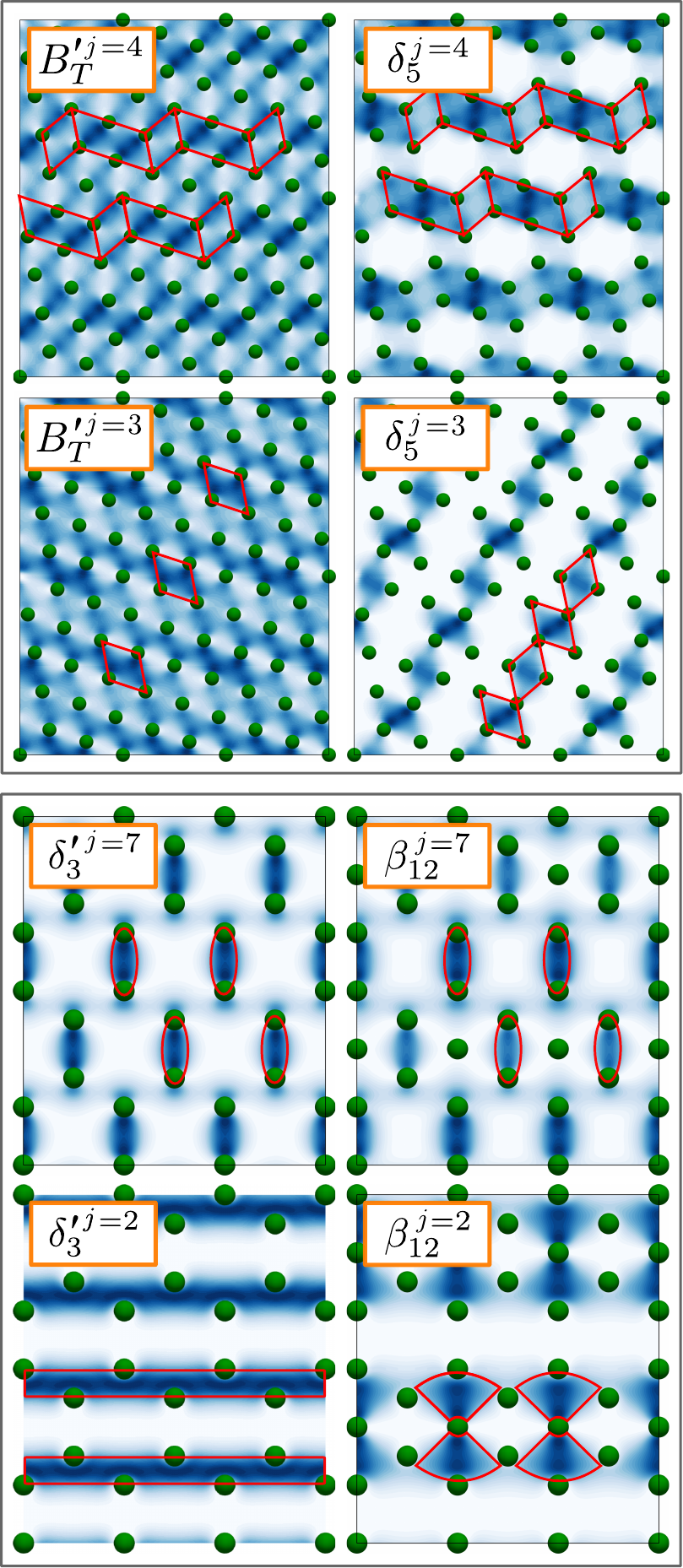}
\caption{\label{fig:PartialDensity_SimilarAndDifferent} Partial densities of ($B'_T\protect\overset{\circ}{\longrightarrow}\delta_5$)
in the top panel, and of ($\delta'_3\protect\overset{\bullet}{\longrightarrow}\beta_{12}$)
in the bottom panel. All states are at $\mbold{k}=(0.10, 0.00, 0.00)$ and $\mbold{k}=(0.12, 0.00, 0.00)$ for
($B'_T\protect\overset{\circ}{\longrightarrow}\delta_5$) and ($\delta'_3\protect\overset{\bullet}{\longrightarrow}\beta_{12}$),
respectively.
}
\end{figure}

Analyzing wavefunctions gives complementary insight with respect to energies. 
The following comparisons are based on the partial densities
$\mathrm{B}^{j\mbold{k}}\equiv |\phi_{j\mbold{k}}(\mbold{r})|^{2}$, where  
$\phi_{j\mbold{k}}(\mbold{r})$ are Kohn-Sham Bloch states and  $\mathrm{B}$ refers to a specific polymorph, i.e., 
$\mathrm{B}=B_{T},\ \delta_{6},\ \delta_{3},\ \alpha_{1},\ \alpha^{\prime},\
\delta_{5},\ \beta_{12},\ \alpha^{\prime}$-Bil.
It is particularly interesting to compare the evolution of weakly dispersing and strongly dispersing states when going from the parent to the child structures. 
To this aim,
we show partial densities in commensurable unit cells, for an arbitrary $\mbold{k}-$point chosen to be close to $\Gamma$,
$\mbold{k}=(0.10,0.00,0.00)$ for the $B'_{T}$ and $\delta_{5}$ cases, and $\mbold{k}=(0.12,0.00,0.00)$ for the $\delta'_{3}$ and $\beta_{12}$.
In the following, 
we will drop the superscript ${\bf k}$. The comparison of the partial densities is shown in Fig. \ref{fig:PartialDensity_SimilarAndDifferent}. 
The top panel compares ($B'_T\overset{\circ}{\longrightarrow}\delta_5$). Whereas a chain-like structure can be detected in all states, the difference between $B_T^{\prime\ j=4}$ and $\delta_5^{j=4}$ is much less pronounced than that between $B_T^{\prime\ j=3}$ and $\delta_5^{j=3}$ where the latter belongs to the weakly dispersing band and shows pronounced localization of charge. 
For ($\delta'_3\overset{\bullet}{\longrightarrow}\beta_{12}$) in the bottom panel of Fig.
\ref{fig:PartialDensity_SimilarAndDifferent} this trend is even more obvious: whereas ($\delta_3^{\prime\ j=7}\leftrightarrow\beta_{12}^{j=7}$) shows close resemblance, the state 
$\beta_{12}^{j=2}$ belonging to the weakly dispersive band breaks the chain-like charge distribution of its parent state $\delta_3^{\prime\ j=2}$ and leads to a periodic localization of charge. This localization of charge is typically found for defects, and we will therefore in the following refer to these states as defect-like, although the corresponding atomic structure would correspond to a very high density of defects. Indeed, as can be seen from both the remaining weak dispersion and weak overlap of charge, the most appropriate picture is that of weakly interacting defects. It is also interesting to note that these observations are similar to those in kagome and pyrochlore lattices, where frustration linked to orbital symmetry leads to reduced hopping and hence, flat bands \cite{Zeng2024}.

The occurrence of the defect-like states is not limited to the polymorphs discussed above, but a similar picture is found in other structures such as $\alpha_{1}$, which is discussed in the following, 
and $\alpha^{\prime}$, whose electronic structure is shown in the 
Appendix, Fig. \ref{fig:OtherDefectLikeStates}, and which can be compared to both $\delta_{3}$ and $B_T$.

\subsubsection*{Defect-like states in larger structures}
\begin{figure}[!ht]
\centering
\includegraphics[width=0.49\textwidth]{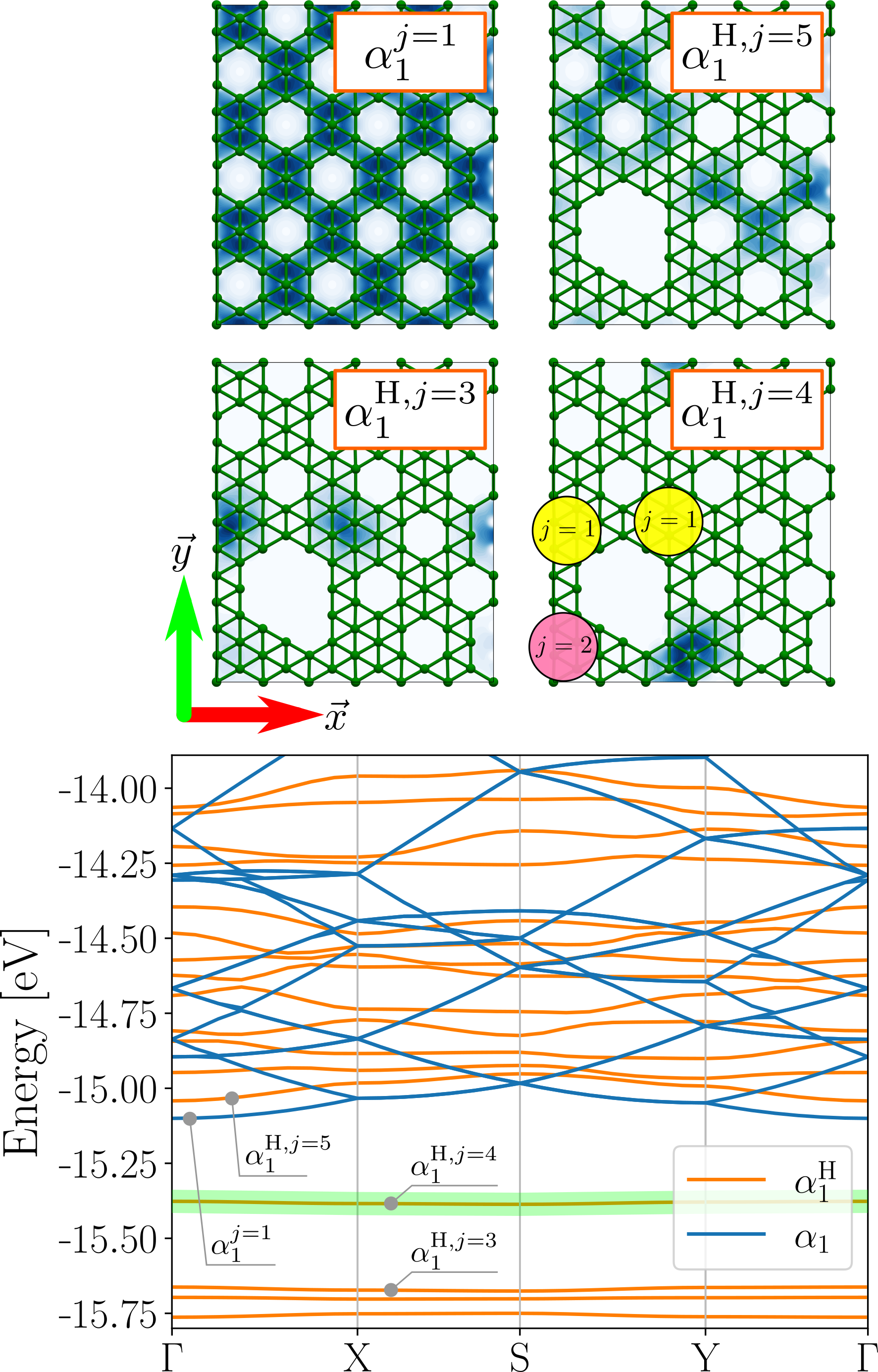}
\caption{\label{fig:alpha1_alpha1H_BandsLow}
Top panel:  In-plane cut of the partial density of the states $\alpha_{1}^{j=1}$, 
$\alpha_{1}^{\mathrm{H},j=3}$, $\alpha_{1}^{\mathrm{H},j=3}$ and 
$\alpha_{1}^{\mathrm{H},j=4}$ at $\Gamma$. Localized states close to the large hole appear in  
$\alpha_{1}^{\mathrm{H},j}$ with $j=1$, 2, 3 and 4; here, we only show $j=3$ and $j=4$. States  
$j=1$ and $j=2$ localize in the zones labeled in yellow and pink, respectively.
Bottom panel: lowest valence bands of $\alpha_{1}$ and 
$\alpha_{1}^{\mathrm{H}}$. The Fermi levels are aligned at 0 eV, though out of range.}
\end{figure}

Using DFT $+$ kinetic Monte Carlo simulations, Park \textit{et. al.} \cite{10.1039/D2NH00518B} observed that liquid borophene with a vacancy ratio
$\eta>\sfrac{1}{8}$ crystallizes under cooling as $\alpha_{1}$, but with large holes with respect to the ideal structure. These holes are limited to a maximum size, and they are protected by a double boron chain that forms around them. 
Here, we created a smaller structure with similar characteristics, namely, a
$\alpha_{1}$ supercell with a large hole protected by a double boron chain. The primitive cell contains 123 atoms and has a vacancy ratio  $\eta=\sfrac{21}{144}$. We refer to this structure
 as $\alpha_{1}^{\mathrm{H}}$. It is shown in the top right panels of 
Fig. \ref{fig:alpha1_alpha1H_BandsLow}.
The lowest bands of $\alpha_{1}^{\mathrm{H}}$ and $\alpha_{1}$ in the commensurable supercell are shown in the bottom panel.

Three aspects are noteworthy: First,
compared to $\alpha_1$, the  width of the valence band of $\alpha_1^H$ increases from 15.1 eV to 15.8 eV. Second,  most of the bands are much less dispersive in 
$\alpha_{1}^{\mathrm{H}}$ than in $\alpha_1$, and the lowest four 
bands are pushed down. Third, starting from $\alpha_{1}^{\mathrm{H},j=5}$, which is found at
15.04 eV at $\Gamma$, the bands of $\alpha_{1}^{\mathrm{H}}$ fall in the energy range of the valence bands of $\alpha_{1}$. In order to characterize the two groups of bands, the top panels of Figure \ref{fig:alpha1_alpha1H_BandsLow} also show the partial 
density of $\alpha_{1}^{\mathrm{H}}$ with $j=3$, 4, 5, together with
$\alpha_{1}^{j=1}$ at $\mbold{k}=(0.0,0.0,0.0)$.
In correspondence with the global flattening of the bands, $\alpha_{1}^{\mathrm{H},j=5}$ is more localized than the perfectly periodic $\alpha_{1}^{j=1}$. Still, there is a clear resemblance, with islands of $\alpha_{1}^{j=1}$-like charge distributed around hexagons in $\alpha_{1}^{\mathrm{H},j=5}$. In this case, one may still recognize a link between parent and child structures. Instead,
the flat band $\alpha_{1}^{\mathrm{H},j=4}$ corresponds to a clearly localized partial density in the atomically denser 
zones next to the hole. The same holds for the lowest states $\alpha_{1}^{\mathrm{H},j=1}$, 
$\alpha_{1}^{\mathrm{H},j=2}$ and $\alpha_{1}^{\mathrm{H},j=3}$, in different sites around 
the hole.
$\alpha_{1}^{\mathrm{H},j=1}$
and $\alpha_{1}^{\mathrm{H},j=3}$ localize in the same zone. These lowest bands can clearly be called localized defect states.

\subsection{Buckled monolayers} \label{section:Results_ElectronicStructure_BuckledMonolayer}

\subsubsection{From flat to buckled structures}
Freestanding monolayers often buckle 
\cite{10.1021/nn302696v,10.1021/acs.jpclett.7b00891,10.1021/jp305545z}. We denote the effect of buckling
with respect to a parent structure with ($\overset{\wedge}{\longrightarrow}$), where the
buckling height is given as the out-of-plane distance ($z$ component in Figure \ref{fig:AtomicStructures_Polymorphs}) from the in-plane position. In particular, we will investigate
the effect of the transitions ($\alpha\overset{\wedge}{\longrightarrow}\alpha^{\prime}$) and
($B_T\overset{\wedge}{\longrightarrow}\delta_6$). Both of them buckle along the
direction of an imaginary phonon mode of the parent structure \cite{PhysRevB.72.045434, 10.1021/jp8052357,
10.1021/nn302696v}. The buckling height is $\pm0.085$ for $\alpha'$ (optimized in the present work) and $0.80\ \mrm{\AA}$ for 
$\delta_6$ (Ref. \cite{10.1126/science.aad1080}). For comparison, we evaluated the band structures, partial densities and Fermi surfaces for different values of the buckling. 
In the case of  ($\alpha\overset{\wedge}{\longrightarrow}\alpha^{\prime}$), 
the atoms in the $2d$ sites are displaced perpendicular to the planes (see 
Table \ref{tab:StructuralParameteresofAtomicPositions}). In the case of ($B_T\overset{\wedge}{\longrightarrow}\delta_6$) the buckling goes together with an in-plane distortion. In order to isolate the out-of-plane effect, we first fix the in-plane distortion 
starting from $B_T$, by matching the projected positions of the children structure ($\delta_6$) . This breaks the hexagonal symmetry, but the effect on the band structure is very mild (see \appendixSI, 
Fig. \ref{fig:BTd6Unbuckled}). We will refer to the flat configuration as $B''_{T}$.
We then study the evolution of the electronic structure with out-of-plane buckling while keeping the in-plane distortion fixed.

\begin{figure*}[!ht]
\centering
\includegraphics[width=0.95\textwidth]{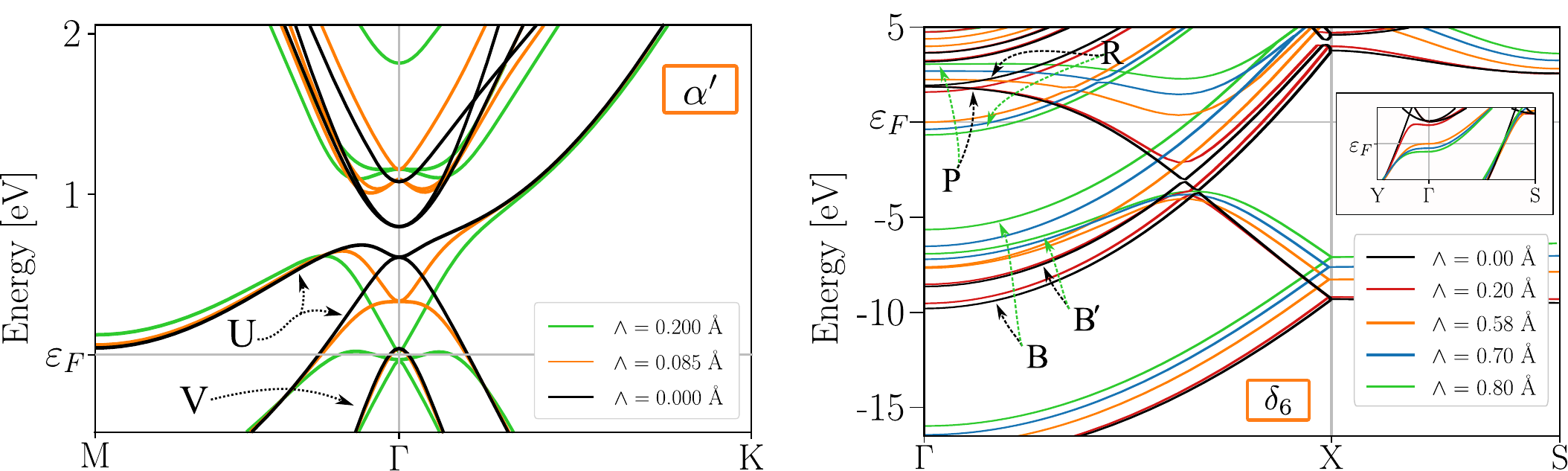}
\caption{\label{fig:BandStructueWithBuckling}
Left: Band structure of borophene 
$(\alpha\overset{\wedge}{\longrightarrow}\alpha^{\prime})$ for different 
buckling heights $\Lambda$. The relaxed $\alpha'$ structure corresponds to $\wedge=0.085$\AA. $\mbold{U}$ labels the two lowest conduction bands degenerate at $\Gamma$ and $\mbold{V}$ the lowest band in frame. Right: $(B''_T\overset{\wedge}{\longrightarrow}\delta_{6})$ for different
buckling heights $\Lambda$. The relaxed $\delta_6$ structure corresponds to $\wedge=0.8$ \AA. 
$\mbold{B}$, $\mbold{B'}$, $\mbold{P}$, and $\mbold{R}$ label specific bands.
Inset in right panel: the band structure of $\delta_6$ close to the Fermi energy for other high-symmetry directions.
}
\end{figure*}

For both ($\alpha\overset{\wedge}{\longrightarrow}\alpha^{\prime}$) and
($B''_T\overset{\wedge}{\longrightarrow}\delta_6$), even for small buckling there is a clearly observable qualitative change of the band structure. In particular, in ($\alpha\overset{\wedge}{\longrightarrow}\alpha^{\prime}$), shown in the left panel of Fig.
\ref{fig:BandStructueWithBuckling}, the lowest conduction bands are pushed down around $\Gamma$, while they remain essentially unchanged further away from $\Gamma$ and towards the borders of the Brillouin zone. This leads to a sharp dip in the conduction band at $\Gamma$ and to a portion of almost flat band around the Fermi level for the highest buckling ($\Lambda=0.2 \text{\AA}$), both stemming from the evolution of the $\mbold{U}$ pair of bands. At the same time,
the next two conduction bands at $\Gamma$ change from parabolic to a mexican-hat like shape. Higher conduction bands are moving upwards.

The band structure of ($B''_T\overset{\wedge}{\longrightarrow}\delta_6$) is shown in the right panel of Fig.
\ref{fig:BandStructueWithBuckling}. Close to $\Gamma$ the two valence bands $\mbold{B}$ and $\mbold{B'}$ move up and exchange their position with increasing buckling, as can be seen by following the bands (indicated with black arrows and green arrows, for zero and $0.8$ \AA{} buckling, respectively). Band $\mbold{B'}$ 
avoids crossing the Fermi energy for buckling $>0.2$ \AA. At the same time, conduction bands $\mbold{R}$ and $\mbold{P}$, which were degenerate at $\Gamma$, split, with the upwards dispersing band $\mbold{R}$ moving below the Fermi energy and the downwards dispersing band $\mbold{P}$ moving upwards. This band, which crosses the Fermi level between $\Gamma$ and $X$ in absence of buckling, remains completely above the Fermi level for the higher bucklings.

\subsubsection{Around the Fermi surface}

\begin{figure}
\centering
\includegraphics[width=0.48\textwidth]{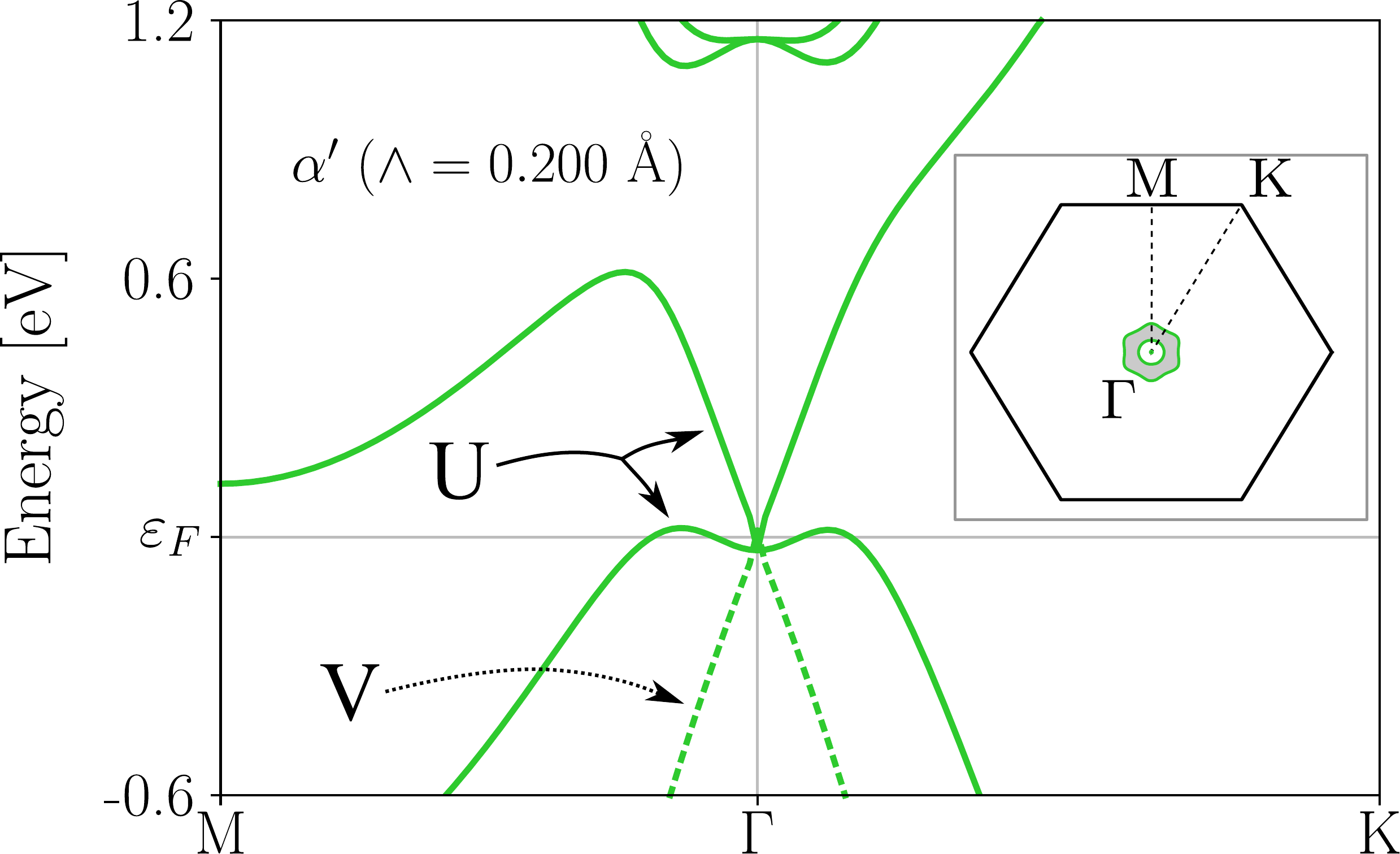}
\caption{\label{fig:BandsWithBucklingSemimetallic}Band structure of $\alpha'$ with $\wedge=0.2\ \mrm{\AA}$. The $\mbold{V}$-band is
is drawn with a dashed line for visual guidance. Inset: Fermi surface with the hole pocket highlighted in grey.}
\end{figure}

The maybe most interesting part of the evolution of the bands in Fig. \ref{fig:BandStructueWithBuckling} occurs close to the Fermi surface. 
Indeed, research has been directed, for instance, to find Dirac cones at the Fermi level, or semi-metallic borophene polymorphs. In particular,
borophene-$P6/mmm$\cite{10.1021/acs.nanolett.5b05292}, $Pmmn$\cite{10.1103/PhysRevLett.112.085502} and
$\chi$-$h_{1}$\cite{10.1021/acs.jpclett.7b00891} have been reported to be semi-metallic. 
Qualitative change of the band structure with buckling is found in the present work for 
$\alpha^{\prime}$. The flattening of the lowest conduction band 
aound $\Gamma$, at the equilibrium value $\Lambda=0.085 \AA$, becomes an inversed mexican hat
for big buckling and deserves closer attention. We can, in fact, push the buckling to the value $\Lambda=0.2 \mathrm{\AA}$, and analyze the results in Fig. \ref{fig:BandsWithBucklingSemimetallic}.

First of all, the system now presents almost a semi-metallic structure with the $\mbold{U}$ and $\mbold{V}$-bands just below
and above the Fermi energy at $\Gamma$, respectively (see \appendixSI, Section\ \ref{sec:S5},
for an analysis of the partial densities for increasing buckling).
This arises from the evolution of the bands with increasing buckling height.
For instance, a Dirac-cone shape has been obtained, with similar mechanism (buckling), and
slightly different unit cell structure ($a=5.046,\ b=5.044,\ \wedge = 0.17\ \text{\AA}$)\cite{10.1021/nn302696v}. In the latter Reference we can see the same
band structure, dispersion, Dirac cone and features at the Fermi energy, that we have evaluated for a slightly bigger buckling. In addition, 
Ref.\cite{10.1021/nn302696v} presents also results for a hybrid exchange-correlation functional (PBE0). The only discrepancy between our results and 
Ref.\cite{10.1021/nn302696v} is coming from an erroneous Fermi energy in their unbuckled $\alpha$ structure, 
which is not compatible with the number of electrons in the unit cell, and the inverse naming of the high-symmetry points $K$ and $M$.
The second important feature given by pushing the buckling to $0.2$ \AA{}, is the creation of a portion of an almost flat band (lowest $\mbold{U}$-band) at the Fermi energy, whose weak inverse mexican hat shape creates a small hole pocket close to $\Gamma$. 
The hole pockets lead to a Fermi surface, shown in the inset of 
Fig.~\ref{fig:BandsWithBucklingSemimetallic}, with 
parallel regions perpendicular to $\Gamma\rightarrow\mathrm{K}$. This indicates 
nesting of the Fermi surface that could be an indicator for charge density waves and for phonon-mediated superconductivity, as it is
inferred in Ref. \cite{10.1063/1.4953775}. 

 In order to investigate further the buckling-induced Fermi-surface nesting, let us consider the polymorph ($B''_T\overset{\wedge}{\longrightarrow}\delta_6$).  Indeed,
Xiao \textit{et. al}, studied the effect of strain and carrier-doping in this polymorph, and found
hole-doping to be beneficial for superconductivity, while electron-doping had the opposite effect
\cite{10.1063/1.4963179}. 
In the following, we complete the work of Xiao \textit{et. al} by a systematic exploration of the effects of buckling and doping, in relation to the evolution of bands.

\begin{figure}
\centering
\includegraphics[width=0.45\textwidth]{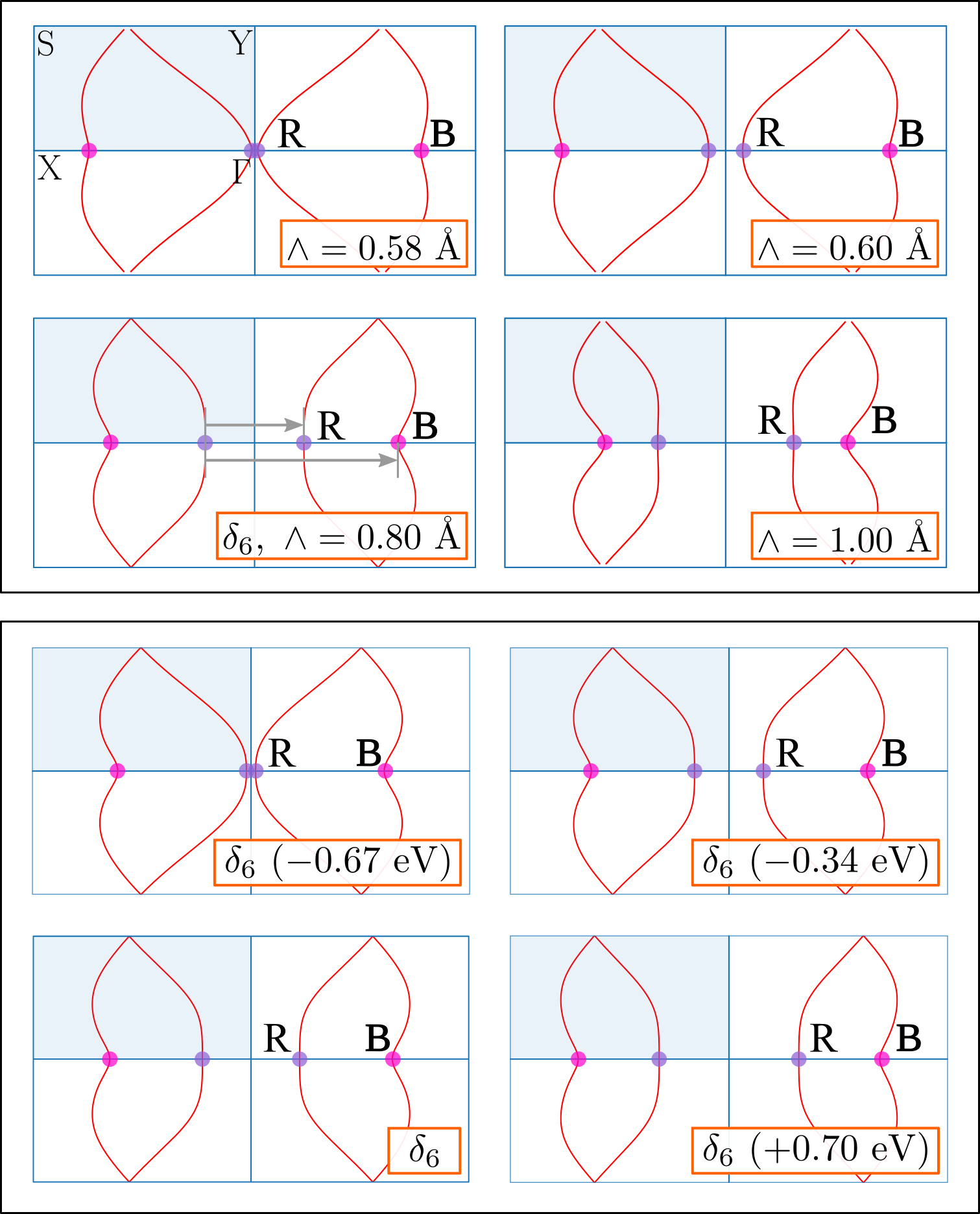}
\caption{\label{fig:g6973}Fermi surface of $\delta_6$ for different values of buckling (top panel) and doping (bottom
panel). The doping results were obtained in the rigid band model. The surfaces closer and further away from the $\Gamma$
point are due to bands $\mbold{R}$ and $\mbold{B}$ defined in the left panel of Fig. \ref{fig:BandStructueWithBuckling}, respectively}
\end{figure}

With increasing buckling, there is a small 
decrease of the band width with and a qualitative change of bands crossing the Fermi level, as can be seen in Fig. \ref{fig:g6973} in the $\Gamma$-X, and close to $\Gamma$ in the  $\Gamma\rightarrow\mathrm{Y}$ and $\Gamma\rightarrow\mathrm{S}$ directions. Most importantly, 
band  $\mbold{P}$ moves upwards, until it doesn't cross the Fermi energy any more when $\wedge\approx 0.50\ \mathrm{\AA}$. At the same time, band $\mbold{R}$ moves downwards and crosses the Fermi energy when $\wedge\approx 0.58\ \mathrm{\AA}$. With further increase in buckling this tendency continues, such that the Fermi-level crossing of the 
$\mbold{R}$ band happens further and further away from $\Gamma$. The inset of the figure also shows that band {\bf R} becomes quite flat in Y-direction with increasing buckling. The second crossing in $\Gamma$-X direction is due to band {\bf B}, with a crossing close to the X point at small buckling that moves closer to $\Gamma$ with increasing buckling. This explains the shape and the evolution of the Fermi surface, shown in Figure
$\ref{fig:g6973}$.
The flattening of the {\bf R}-band in $\Gamma$-Y direction is responsible for the almost parallel portions of the Fermi surface that develop with increasing buckling. 
Such a Fermi surface nesting 
has been identified as the cause of Kohn anomalies in $\delta_6$, and therefore as the probable origin of high electron-phonon coupling at the
corresponding $\mbold{q}$ \cite{10.1021/acs.nanolett.6b00070,10.1063/1.4963179}.
More precisely, the two Kohn anomalies, related to negative phonon frequencies in Fig. 3 of
Ref.\cite{10.1021/acs.nanolett.6b00070}, correspond to the two momenta depicted as grey arrows in 
the top panel of Fig.\ref{fig:g6973}. In other words, buckling is expected to be at the origin of significant differences between the at first sight quite similar structures $B_T$ and $\delta_6$, due to
the nesting
contribution to the electron-phonon coupling \cite{10.1021/acs.nanolett.6b00070,
10.1103/RevModPhys.73.515}.

Varying the buckling of $\delta_{6}$ further continues to change the Fermi surface as depicted in the top panel of 
Fig. \ref{fig:g6973}. First, the nesting vector governed by the $\mbold{R}$ band increases with buckling. Second, at the same time the section of almost parallel Fermi surface becomes more extended, which enhances the effect of nesting.
Since the position of the {\bf R}-band is responsible for this change, the self-doping model suggests that a similar effect could also be obtained by doping instead of buckling. This is indeed the case, as shown in the bottom panel of Fig. \ref{fig:g6973} obtained in the rigid band model. It is especially interesting, since doping can be controlled experimentally, through a substrate or contacts, more easily than buckling in a non-polar material.

\subsection{$\alpha^{\prime}$-bilayer}\label{section:Results_ElectronicStructure_Bilayer}

Finally, let us compare the $\alpha^{\prime}$-bilayer (AA-stacked $\alpha^{\prime}$-bilayer) to
the $\alpha^{\prime}$ monolayer.
As discussed earlier and as shown in Fig. \ref{fig:AtomicStructures_Polymorphs}, $\alpha^{\prime}$ is slightly buckled. 
In the bilayer this distortion has a preferential direction, with a larger buckling towards the second monolayer. This leads to covalent
bonding between its 6-coordinated atoms, which drags these atoms towards the central plane of the structure \cite{10.1038/s41563-021-01084-2}. It is in contrast to the case of flat monolayers, where the lowest out-of-plane states are decoupled from the in-plane states, with a partial density that  simply localizes above and below the atomic positions (see \appendixSI, Fig. \ref{fig:BoropheneFlatPartialDensityOutOfPlane}). 

\begin{figure*}
\centering
\includegraphics[width=0.98\textwidth]{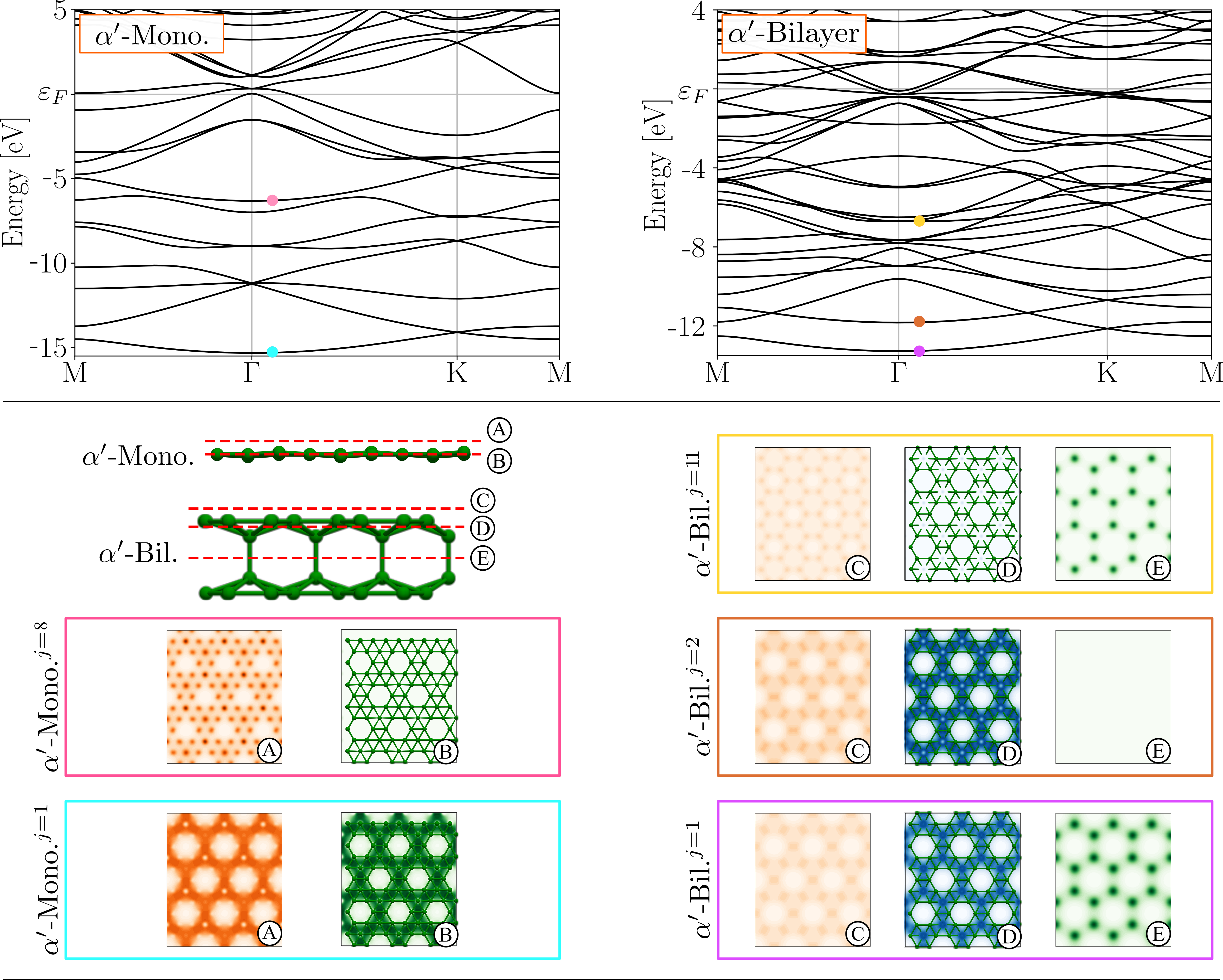}
\caption{\label{fig:BondingAndtibonding_Bilayer}
Band structure of single (top left) and double layer (top right) $\alpha^{\prime}$. The partial density of highlighted states
is shown in the panels below, all calculated at $\mbold{k}=(0.05,0.03,0.00)$. The different cuts are shown in dashed lines in the
atomic structures: A and C are 0.4 $\mrm{\AA}$ above the monolayer and bilayer, respectively.
}
\end{figure*}

Fig. \ref{fig:BondingAndtibonding_Bilayer} shows the band structures of the single and bilayer structures, as well as the partial densities
for selected states at $\mbold{k}=(0.05,0.03,0.00)$.
When the layers are brought together, the two single layer states evolve into the lowest bilayer state, which has interlayer-bonding character. In correspondence, there is a second state with similar dispersion but higher energy, which is of anti-bonding character. 
This can be seen clearly in the partial densities, which has significant contribution in the central plane for $\alpha'$-Bil.$^{j=1}$, but not for $\alpha'$-Bil.$^{j=2}$, both resembling $\alpha'$-Mono.$^{j=1}$.

It is also interesting to follow the evolution of $p_z$-states of the monolayer. The partial charge of these states is localized above and below the atoms. In the bilayer, the first state that clearly localizes charge out of plane is $\alpha'$-Bil.$^{j=11}$ (highlighted in yellow in Fig. \ref{fig:BondingAndtibonding_Bilayer}). In this state, charge is attracted towards the central atom of the hexagons, localizing mostly between the planes.
These findings might help to interpret future experiments. 
Indeed, while these systems start to be explored in depth theoretically, for example, bilayer $\delta_6$ \cite{Zhong2024}, 
the synthesis of bilayer borophene is still in its early stages, and only a small number of structures could so far be realized experimentally.

\section{Conclusions}
\label{sec:conclusions}

In this article, we carried out a detailed analysis of the electronic structure of a wide range of borophene 
polymorphs. We showed how the strong similarity among the different polymorphs can be exploited to partially 
describe the electronic properties of borophene, starting from the knowledge of few parent-like structures, 
using the self-doping picture. The limitations of this approach are highlighted, thanks to the analysis not 
only of the band-structure, but also of the wave functions describing the single particle states, via an 
illustrative depiction of the partial electron densities. This visual analysis permit us, in particular, 
to address the description of low-dispersive polymorph-specific states, that drags the electron cloud towards the defect,
and are objects of recent research in different fields
including quantum Hall physics, superconductivity and photonics. 
This strategy revealed succesfull also to tackle the analysis of a large borophene structure, with big holes as
defect and containing more than a hundred atoms in the unit cell.

In addition to the flat monolayers, we also investigated the bilayer $\alpha'$-Bil, as well as
two buckled structures:  $\alpha'$ and $\delta_{6}$. The bilayer case confirmed the possibility 
of describing some of the features of the child structure, from the knowledge of the parent, and the 
analysis of the partial density reveals the bonding/anti-bonding character to the bilayer states. 
Similarly to the flat monolayers case, even for the bilayer some states are specific to the child
structure, but rather than reveals as low-dispersive states, localized around the defect, the analysis
of the partial density show the out-of plane character of these states, and, consequently the covalent 
nature of the bilayer (also) in the z-direction.   
Finally, the buckled structure have also been investigated into details, and in particular at the Fermi
surface. It is, in fact, particularly interesting to analyze the buckling as a tuning parameter that 
permits to activate exotic band-structure effects, like semimetallicity and Dirac-cones, or the nesting 
of the Fermi surface, linked to known Kohn anomalies, and a precursor for superconductivity. 

\section{Acknowledgments}
This project has received funding from the European
Union’s Horizon 2020 research and innovation programme
under grant agreement no. 800945 - NUMERICS - H2020-MSCA-COFUND-2017.
Computational time was
granted by GENCI (Project No. 544).

\section{Computational details}
\label{sec:compdet}
We used the atomic structures for polymorphs $\delta_6$, $\delta_3$, $\beta_{12}$,
$\alpha_1$ and $\delta_5$ as obtained from Refs. \cite{10.1021/jp066719w, 10.1021/acs.jpclett.7b00891,
10.1021/nn302696v, 10.1126/science.aad1080, 10.1103/PhysRevLett.125.116802, 10.1038/s41563-021-01084-2}
without any further optimization. Instead, to compare $\alpha^{\prime}$-Monolayer with
$\alpha^{\prime}$-Bilayer, we relaxed the atomic positions and lattice parameters
(within a threshold on the maximal force of 0.01 eV/\AA), starting out from the bilayer structure in Ref.
\cite{10.1038/s41563-021-01084-2}. The relaxation of the structures was performed using the optimized
norm-conserving Vanderbilt pseudopotentials (ONCVPSP 3.2.3.1), evaluating the exchange-correlation potential
within the generalized gradient approximation (GGA) using the Perdew-Burke-Ernzerhof (PBE) functional
\cite{10.1103/PhysRevLett.77.3865, 10.1103/PhysRevB.88.085117}. The model system $B_T$ was manually
created taking a bond distance of 1.68 $\mathrm{\AA}$, based on the shortest bond in $\delta_6$. The
symmetry of the polymorphs was determined using FINDSYM \cite{10.1107/S0021889804031528, findsymOnline}.

The ground state and band structures were performed using the Hartwigsen-Goedecker-Hutter
(HGH) pseudopotential within the local density approximation (LDA) \cite{10.1103/PhysRev.140.A1133,
10.1103/PhysRevB.58.3641}, finding minimal differences with respect to those obtained with GGA-PBE
(see, for example, Fig. \ref{fig:Balpha_Bands_LDA_PBE} in \appendixSI - here only LDA
results are included). In the framework of the density functional theory (DFT) we used the Abinit
package \cite{10.1103/PhysRev.136.B864,
https://doi.org/10.1016/j.cpc.2019.107042}. We converged the ground state with respect to the
total energy within a threshold of 0.5 meV/atom. Moreover, we converged the interlayer vacuum distance to
19.05 $\mathrm{\AA}$ (21.17 $\mathrm{\AA}$ exceptionally for $\alpha^{\prime}$-bilayer). For all polymorphs we
used a cutoff energy of 75 Hartree ($\sim 2041$ eV) and the $\mbold{k}$-meshes for the self-consistent calculation
of the KS density for $\delta_6$, $\delta_3$, $\beta_{12}$, $\alpha_1$, $\delta_5$, $\alpha^{\prime}$ and
$\alpha^{\prime}$-bilayer are, respectively: $38\times 38\times 1$, $40\times 40\times 1$,
$32\times32\times 1$, $28\times 28\times 1$, $30\times 30\times 1$, $34\times 34\times 1$ and
$40\times 40\times 1$ \cite{10.1103/PhysRevB.13.5188}. For $\alpha_1^H$ we computed the ground state using a 
$12\times10\times1$ $\mbold{k}$-mesh. All polymorphs here presented are metallic, and
we used a 0.01 Hartree temperature smearing.

\bibliography{apssamp}
\clearpage
\onecolumngrid
\section{\appendixSI}
\label{sec:supp}
\beginsupplement

\subsection{$B_{T}$ atomic structure}

\begin{figure*}[!ht]
\centering
\includegraphics[width=0.78\textwidth]{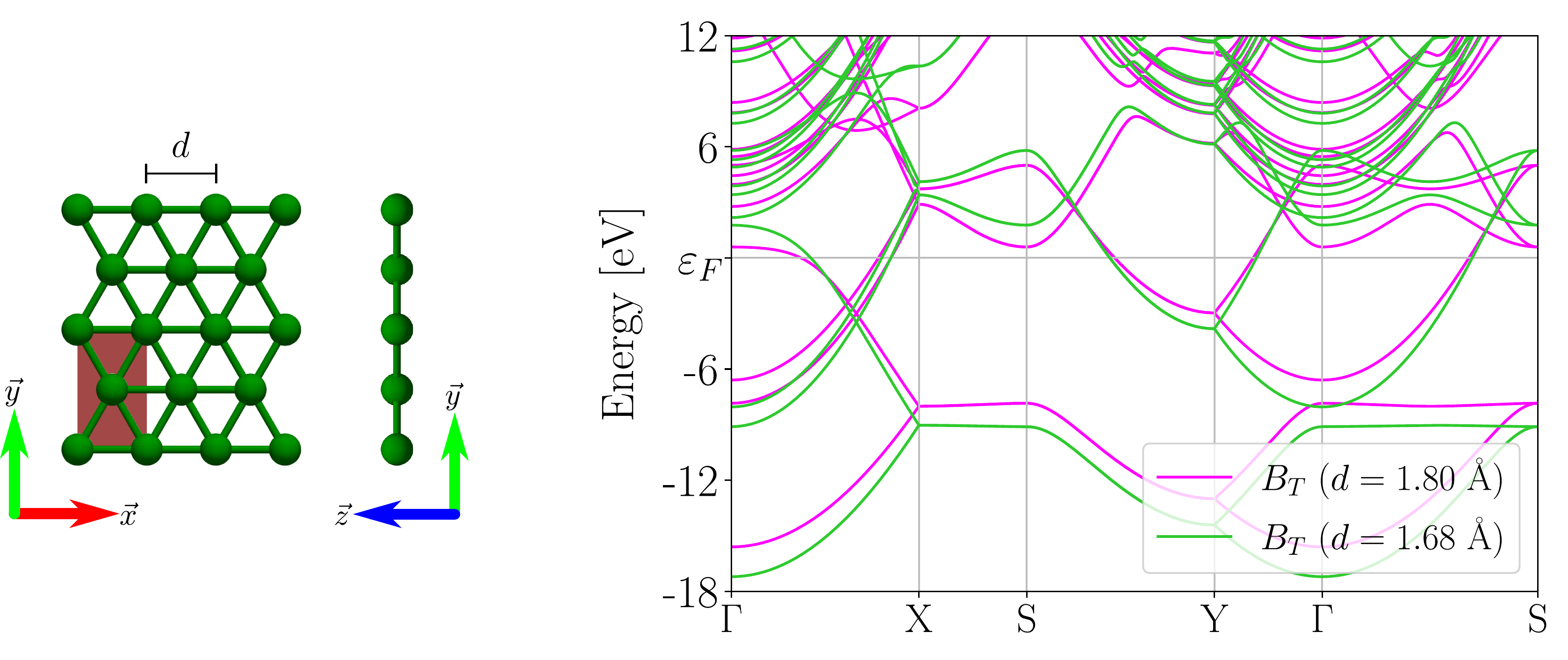}
\caption{\label{fig:BT_diffBonds_WStructure}Electronic band structure of the model system $B_{T}$ with two different bond distances.}
\end{figure*}

We built this structure based on the bond distances of $\delta_6$ (1.68 and 1.80 $\mathrm{\AA}$).
The different choice affects, in particular, the bandwidth of the electronic structure (Fig. \ref{fig:BT_diffBonds_WStructure}).
However, the overall shape of the individual bands is maintained as well as their crossings of the Fermi energy.

\subsection{$\alpha^{\prime}$ electronic structure}

\begin{figure*}[!ht]
\centering
\includegraphics[width=0.70\textwidth]{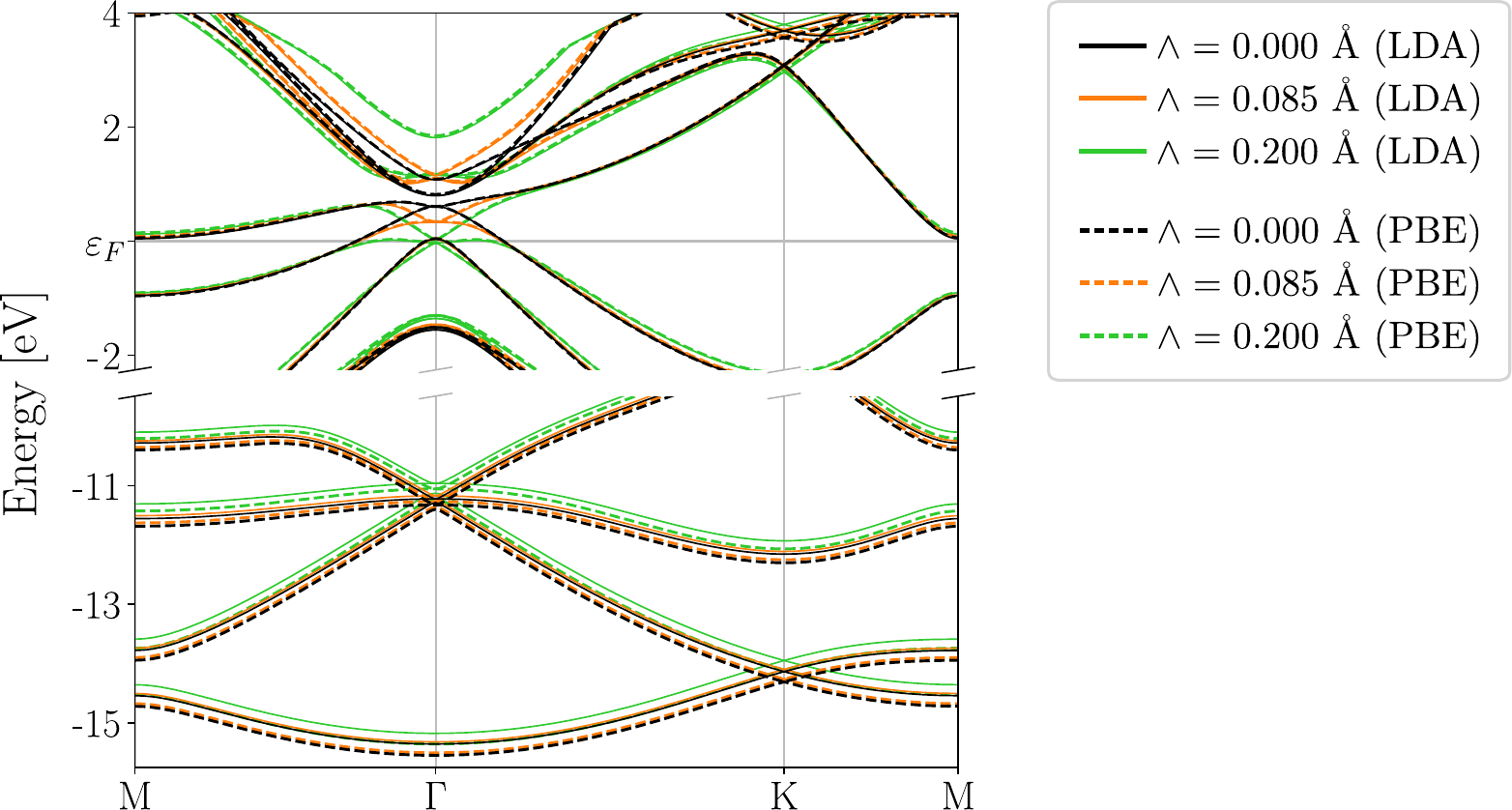}
\caption{\label{fig:Balpha_Bands_LDA_PBE}
Electronic band structure of $\alpha^{\prime}$ calculated with different buckling heights with the LDA and
PBE. Overall, the characteristics of the band spectra for the different buckling configurations are maintained
by using both functionals. The major changes occur upon buckling at $\Gamma$ and near the Fermi energy, these are
discussed in Section \ref{section:Results_ElectronicStructure_BuckledMonolayer}.}
\end{figure*}

The polymorph $\alpha^{\prime}$ can be found in the literature with different buckling heights
\cite{10.1021/nn302696v, 10.1103/PhysRevB.93.165434}. Here we show that different values have a small influence in the
band width and negligible differences in the shape of the valence bands (Fig. \ref{fig:Balpha_Bands_LDA_PBE}).
We also use this example to show the negligible difference at the level of the electronic structure that we obtain
from using different functionals, in this case, LDA and PBE.

\subsection{Other defect-like states}
\begin{figure*}[!ht]
\centering
\includegraphics[width=0.95\textwidth]{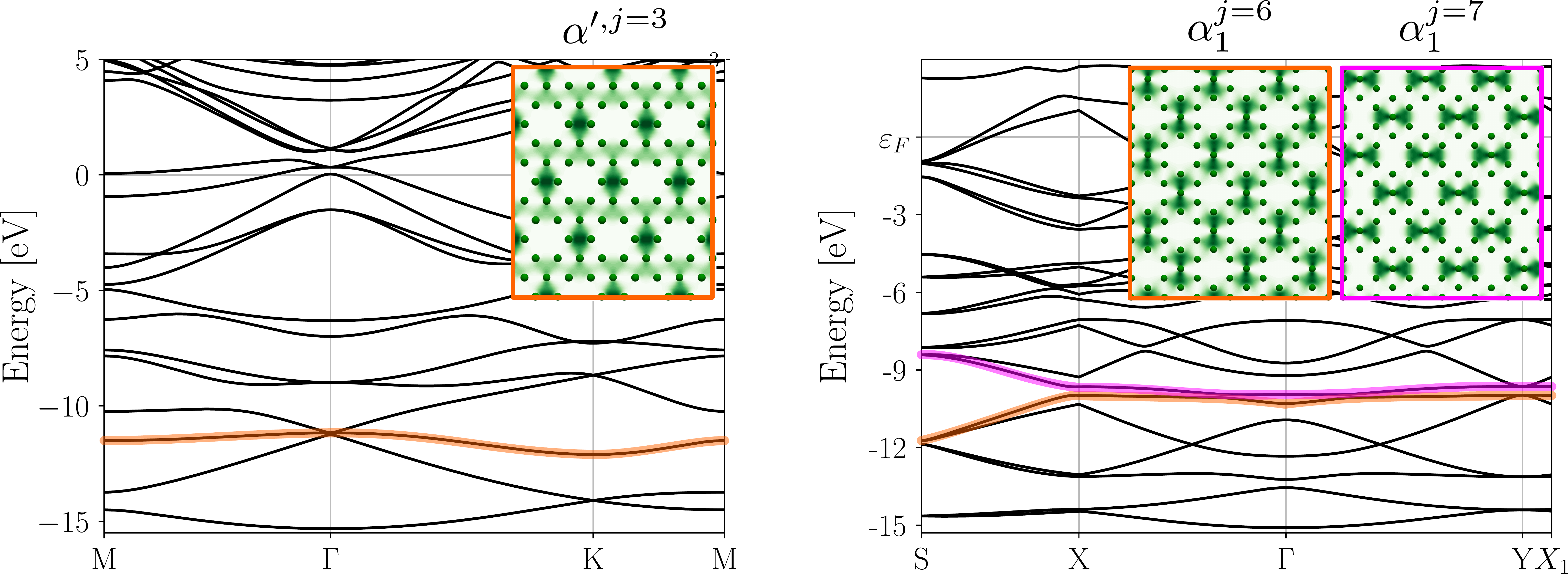}
\caption{\label{fig:OtherDefectLikeStates}Electronic band structure of $\alpha^{\prime}$ and $\alpha_{1}$ highlighting the low-dispersive
defect-like states $\alpha^{\prime,j=3}$, and $\alpha_{1}^{j=6}$ and $\alpha_{1}^{j=7}$. For each case we have added a cut in-plane
of the partial density (insets), showing the localization of the partial charge.}
\end{figure*}

This corresponds to Fig. \ref{fig:OtherDefectLikeStates}.

\subsection{$B_{T}$ and $B''_{T}$}

\begin{figure*}[!ht]
\centering
\includegraphics[width=0.48\textwidth]{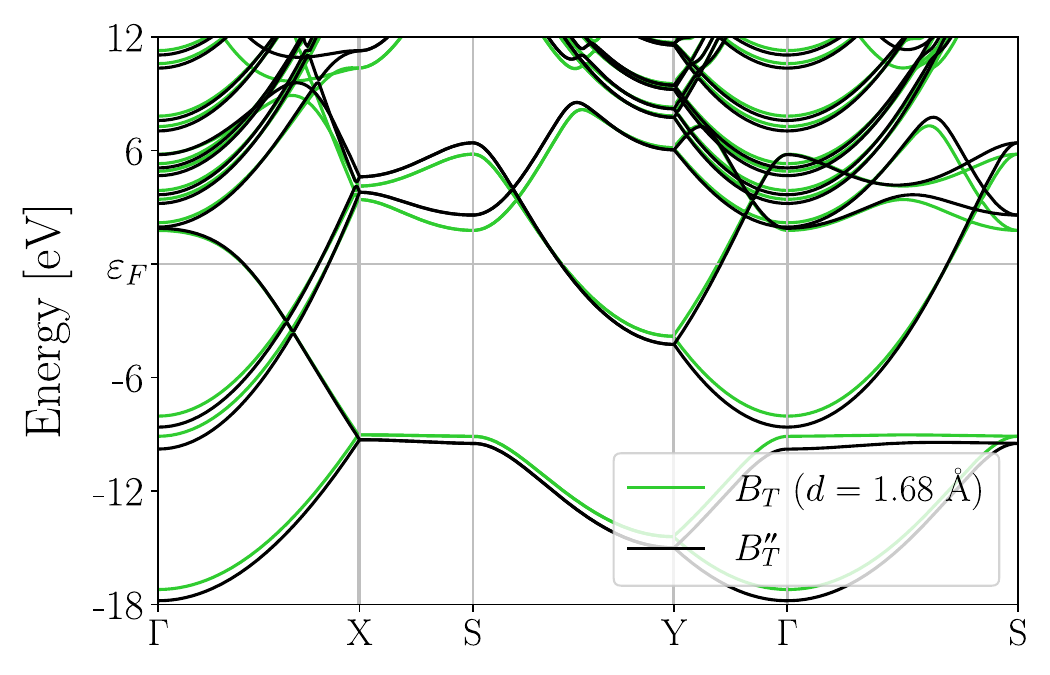}
\caption{\label{fig:BTd6Unbuckled}Electronic structure of model system $B_{T}$ (perfectly regular and planar with bond
distance 1.68 $\mrm{\AA}$) with hexagonal symmetry, and unbuckled $\delta_{6}$, also called $B''_{T}$, with tetragonal
symmetry. Despite the in-plane distortion of the systems compared to each other, the electronic structures are nearly
identical.}
\end{figure*}

This corresponds to Fig. \ref{fig:BTd6Unbuckled}.

\subsection{Partial density of selected states of $\alpha'$ with different buckling heights}
\label{sec:S5}

We computed the partial density of selected states for $\alpha$ and $\alpha'$ with different buckling heights.
A naive plotting of the band structure will break the degeneracy of the $\mbold{U}$-bands and will open a small
gap at $\Gamma$ with $\wedge=0.2\ \mrm{\AA}$. However, in order to determine the correct order of these states
we can follow their evolution upon buckling.

\begin{figure*}[!ht]
\centering
\includegraphics[width=1\textwidth]{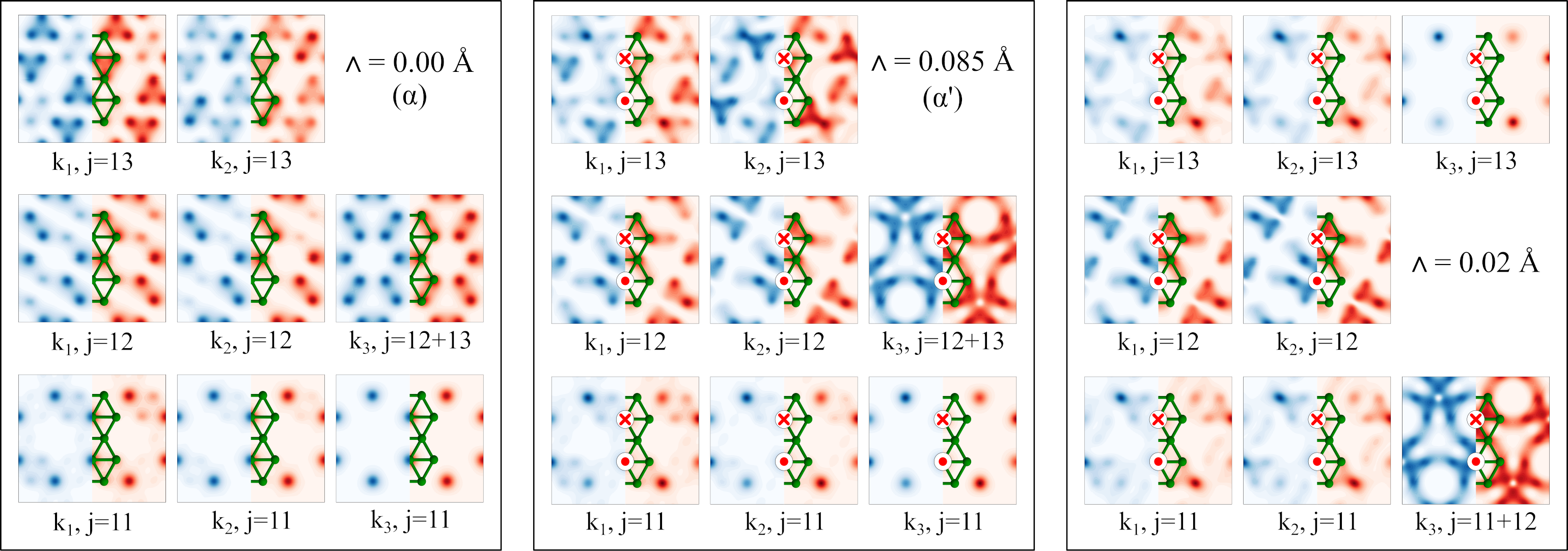}
\caption{\label{fig:DiracConeStates}Partial density cuts of selected states of $\alpha$ and $\alpha'$ with different buckling
heights. The selected states are $j=11$, 12 and 13, which corresponds to the $\mbold{U}$ and $\mbold{V}$-bands in the left panel
of Fig. \ref{fig:BandStructueWithBuckling}. For all cases, $\mbold{k}_{1}=(0.07,0,0)$, $\mbold{k}_{2}=(0.04,0,0)$ and
$\mbold{k}_{3}=\Gamma$. The partial density of these states is mostly out of plane, so the cuts in blue and red are above and below
the mid-plane of the atomic structure, respectively. The atom buckling upwards in $\alpha'$ is shown with a red dot
(\textcolor{red}{$\bullet$}), while the atom downwards with a red cross (\pmb{\textcolor{red}{${\times}$}}).}
\end{figure*}

At a first stage (see left panel of Fig. \ref{fig:BandStructueWithBuckling}), in $\alpha$ ($\wedge=0.00\ \mrm{\AA}$),
the $\mbold{U}$ bands ($j=12$ and $j=13$) are degenerate at $\Gamma$ and clearly separated from the $\mbold{V}$ band ($j=11$).
Upon small buckling ($\wedge=0.085\ \mrm{\AA}$), the $\mbold{U}$ bands near $\Gamma$ move downwards. This change shows
in the partial density cuts of Fig. \ref{fig:DiracConeStates} as a localization toward the atoms out of plane
(see left and central panels for comparison). The localization is visible from the intensity of the colors, and,
for example, it creates the rings structure above and below the plane in $\alpha'(\wedge=0.085\ \mrm{\AA})^{j=12+13}$.

At higher buckling, $\wedge=0.2\ \mrm{\AA}$ (right panel of Fig. \ref{fig:DiracConeStates}), the selected states remain
similar for $\mbold{k}_{1}=(0.07,0,0)$ and $\mbold{k}_{2}=(0.04,0,0)$ to those in the previous panels. It is clear even from the
band structure of Fig. \ref{fig:BandsWithBucklingSemimetallic} that the $\mbold{U}$-bands correspond to $j=12$ and $j=13$,
and the $\mbold{V}$-band to  $j=11$. At $\mbold{k}_{3}=\Gamma$ the index of the bands change. Now, the lowest
bands $j=11$ and $j=12$ are degenerate below the Fermi energy. However, the partial density shows that
$\alpha'(\wedge=0.2\ \mrm{\AA})^{j=11+12}_{\mbold{k}_{3}}$ corresponds to the $\mbold{V}$-band while
$\alpha'(\wedge=0.2\ \mrm{\AA})^{j=13}_{\mbold{k}_{3}}$ to the $\mbold{U}$-band.

\begin{figure}[!ht]
\centering
\includegraphics[width=0.48\textwidth]{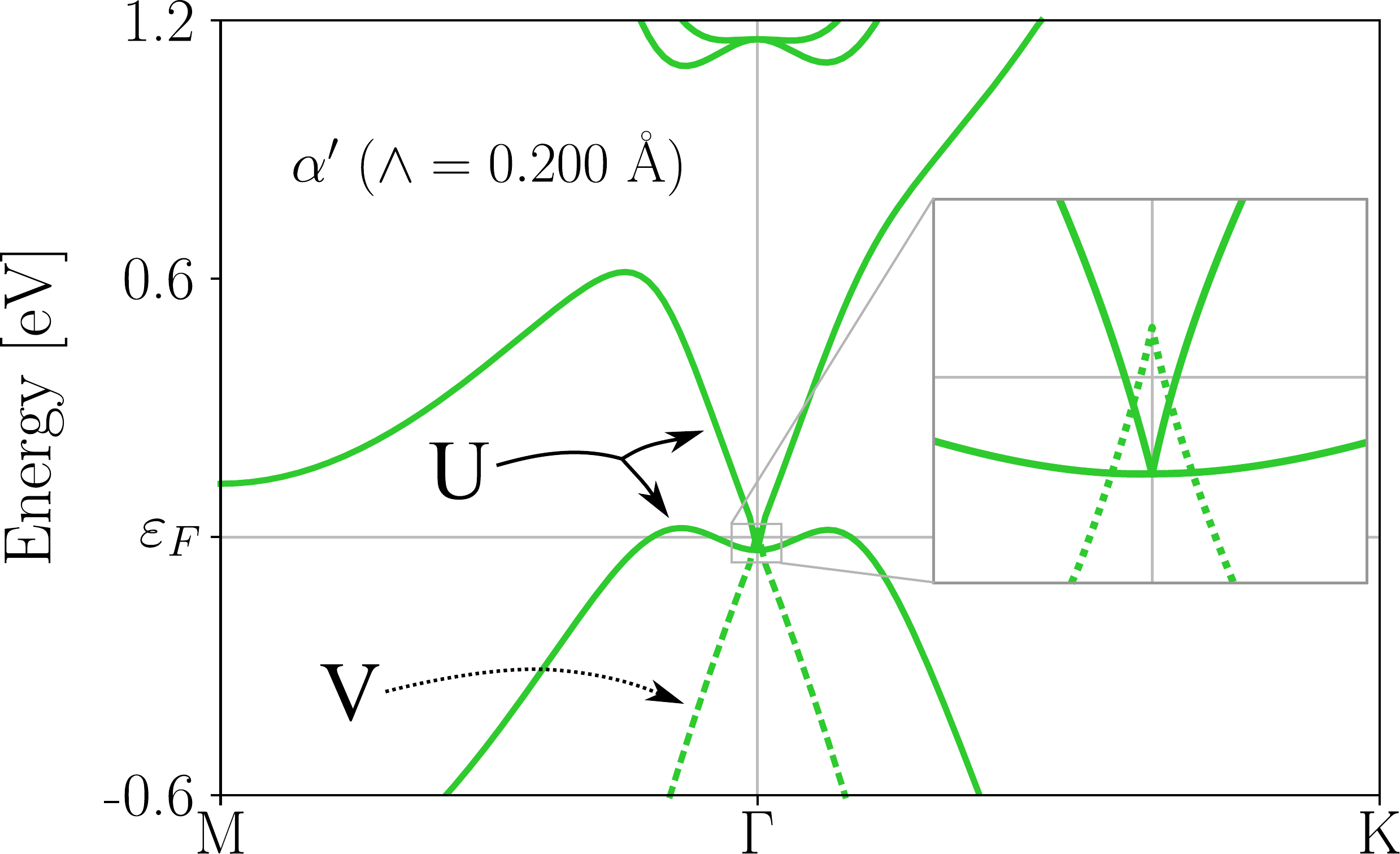}
\caption{\label{fig:BandsWithBucklingSemimetallic_insetb}Band structure of $\alpha'$ with $\wedge=0.2\ \mrm{\AA}$. The $\mbold{V}$-band is
is drawn with a dashed line for visual guidance. Inset: enlarged view of the most probable band structure configuration at $\Gamma$ near the Fermi level.}
\end{figure}

In summary, upon increasing buckling, the $\mbold{U}$-bands near $\Gamma$ are systematically pushed down in energy, and at high
buckling, the $\mbold{U}$-band moves below the Fermi level. This mechanism then creates the band structure as depicted
in Fig. \ref{fig:BandsWithBucklingSemimetallic_insetb}, and its inset.

\subsection{States out of plane of flat monolayers}

\begin{figure*}[!ht]
\centering
\includegraphics[width=0.65\textwidth]{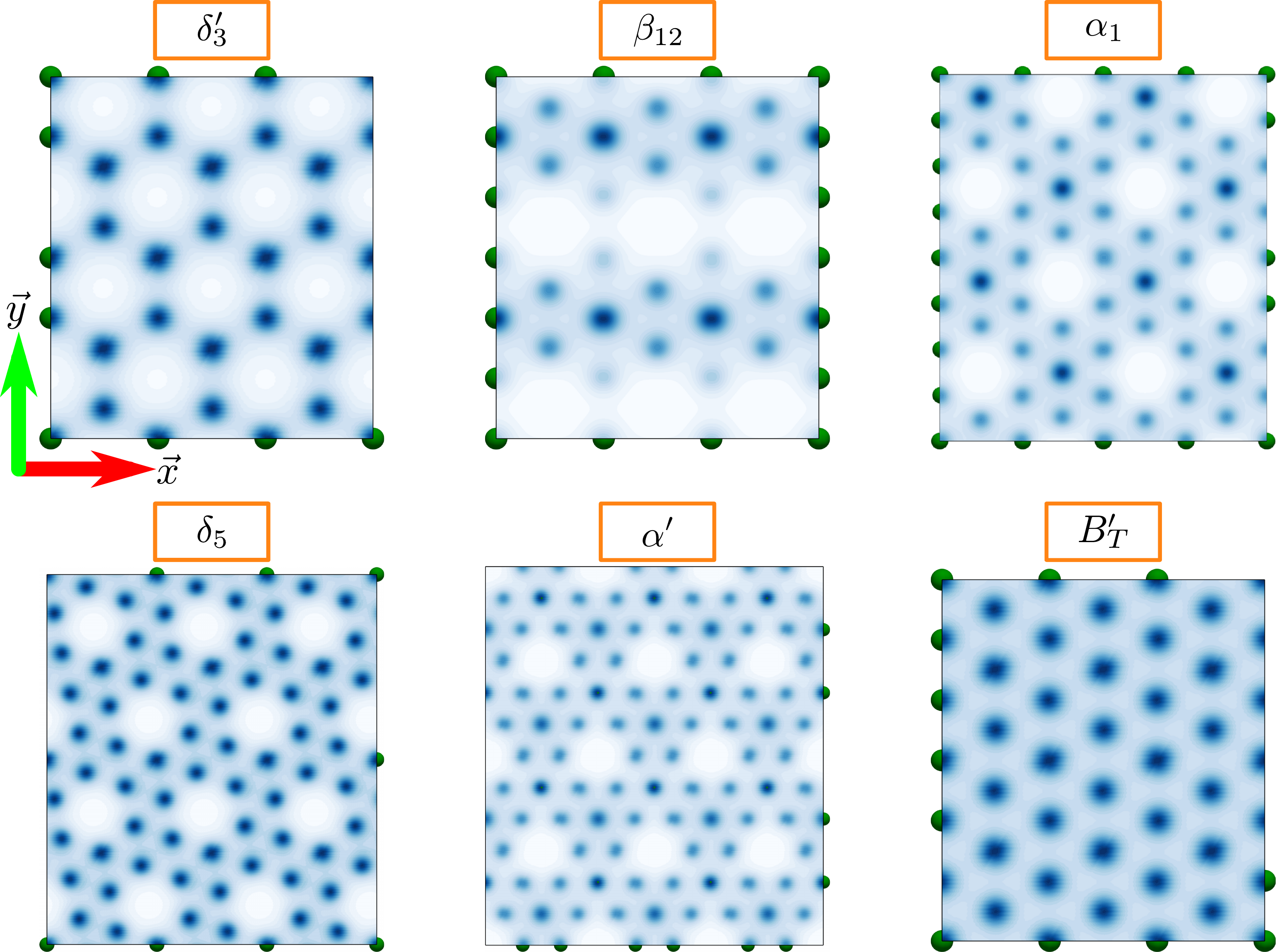}
\caption{\label{fig:BoropheneFlatPartialDensityOutOfPlane}Partial density of the lowest state out-of-plane for each of the
flat polymorphs near. The plots here presented were computed at $\mbold{k}=(0.12, 0.00, 0.00)$ for $\delta'_3$ and $\beta_{12}$,
$(0.05, 0.00, 0.00)$ for $\alpha_1$, $(0.10, 0.00, 0.00)$ for $\delta_5$ and $B'_T$, and $(0.05, 0.03, 0.00)$
for $\alpha^{\prime}$.}
\end{figure*}

This corresponds to Fig. \ref{fig:BoropheneFlatPartialDensityOutOfPlane}.

\end{document}